**Title**

Stacking faults enabled second harmonic generation in centrosymmetric van der Waals RhI$_3$

**Authors**

Yue Liu[1†], Wen He[1†,*], Bing-Ze Wu[4†], Feng-Yuan Xuan[5], Yu-Qiang Fang[3*], Zheng-Bo Zhong[6], Jie-Rui Fu[1], Jia-Peng Wang[1], Zhi-Peng Li[6], Jin-Zhong Wang[1], Ming-Guang Yao[4], Fu-Qiang Huang[3], Liang Zhen[1,2], Yang Li[1*], Cheng-Yan Xu[2*]

**Affiliations**

[1]School of Materials Science and Engineering, Harbin Institute of Technology, Harbin 150001, China.

[2]Sauvage Laboratory for Smart Materials, School of Materials Science and Engineering, Harbin Institute of Technology (Shenzhen), Shenzhen 518055, China.

[3]State Key Laboratory of High-Performance Ceramics and Superfine Microstructure, Shanghai Institute of Ceramics Chinese Academy of Sciences, Shanghai 200050, China.

[4]State Key Laboratory of Superhard Materials, College of Physics, Jilin University, Changchun 130012, China.

[5]Suzhou Laboratory, Suzhou, 215123, China.

[6]School of Chemistry and Chemical Engineering, Shanghai Jiao Tong University, Shanghai 200240, China.

[†]These authors contributed equally to this work
*Corresponding authors. Email: hewenmse@hit.edu.cn (W.H.); fangyuqiang@mail.sic.ac.cn (Y.Q.F.); liyang2018@hit.edu.cn (Y.L.); cy_xu@hit.edu.cn (C.Y.X.)

**Abstract**

Second harmonic generation (SHG) in van der Waals (vdWs) materials has garnered significant attention due to its potential for integrated nonlinear optical and optoelectronic applications. Stacking faults in vdWs materials, a typical kind of planar defect, can introduce a new degree of freedom to modulate the crystal symmetry and resultant SHG response, however, the physical origin and tunability of stacking-fault-governed SHG in vdWs materials remain unclear. Here, taking the intrinsically centrosymmetric vdWs RhI$_3$ as an example, we theoretically reveal the origin of stacking-fault-governed SHG response, where the SHG response comes from the energetically favorable $A\overline{C}$ stacking fault of which the electrical transitions along the high symmetry paths $\Gamma - M$ and $\Gamma - K$ in the Brillion zone play the dominant role at 810 nm. Such stacking-fault-governed SHG response is further confirmed *via* structural characterizations and SHG measurements. Furthermore, by applying hydrostatic pressure on RhI$_3$, the correlation between structural evolution and SHG response is revealed with SHG enhancement up to 6.9 times, where the decreased electronic transition energies and huger momentum matrix elements due to the stronger interlayer interactions upon compression magnify the SHG susceptibility. This study develops a promising foundation based on strategically designed stacking faults for pioneering new avenues in nonlinear nano-optics.

**Introduction**

Symmetry breaking plays a vital role in the fascinating physical phenomena that have been observed in van der Waals (vdWs) materials[1], including Hall effects[2,3], valley-contrasting physics[4], spin-orbit physics[5,6], and nonlinear optics[7–9]. Various approaches such as electric fields[10], strain[11], doping[12], and defects[13] can effectively manipulate the inversion symmetry. Especially, the intrinsic defects, including point defects, line defects, and planar defects created during the growth or preparation processes, are inevitable in vdWs materials and strongly influence their electronic and symmetry structure[14,15]. Notably, stacking faults, a typical kind of planar defect, which are defined as a local deviation from one of the close-packed stacking orders to the other one, hold great potential to determine the space group and introduce a new degree of freedom for the crystal symmetry modulation[16–18].

Given that the stacking faults can induce inversion symmetry breaking, modulating SHG response through strategically designed stacking faults in vdWs materials becomes a feasible approach. For example, the homostructure h-BN with an $AB$ stacking fault exhibits enhanced

SHG than normal AA′ homostructure due to the symmetry breaking despite having the identical layer number[19]. The SHG signal can emerge in the centrosymmetry AB stacking GeSe, resulting from the local AA′ stacking fault[20]. Such stacking faults in vdWs materials, arising from weak interlayer coupling, might undergo unexpected changes under external conditions. Especially external stimuli like high pressure, which can significantly induce interlayer compression to enhance the interlayer coupling[21,22]. However, the physical origin of stacking-fault-induced SHG response and its tunability by external stimulus are still ambiguous, which deserve to be further investigated.

As a member of vdWs transition-metal trihalides $MX_3$ (M = Cr, Rh, and Ru; X = Cl, Br, and I), $RhI_3$ exhibits intrinsically centrosymmetry and low formation energy of stacking faults, making it appreciable to study how stacking faults break inversion symmetry, the origin of SHG response and its pressure-driven tunability. In this work, taking centrosymmetry $RhI_3$ as an example, we theoretically and experimentally investigated possible stacking faults and their effects on SHG, where the type of stacking faults responsible for SHG response and the origin of SHG have been investigated comprehensively. Moreover, by applying hydrostatic pressure on vdWs $RhI_3$ with stacking faults, a significant pressure-driven enhancement of SHG response with the factor up to 6.9 and its microscope mechanism has been revealed by the first principle calculations. We elucidate the structure and origin of the SHG response in $RhI_3$, the enhancement effects under pressure and its physical mechanism *via* experiments and theoretical calculations. This work provides an important insight into the generation and tunability of the SHG response in 2D materials and develops a promising foundation for pioneering new avenues in nonlinear nano-optics.

# Results

## Calculated stacking fault structures and their SHG susceptibility.

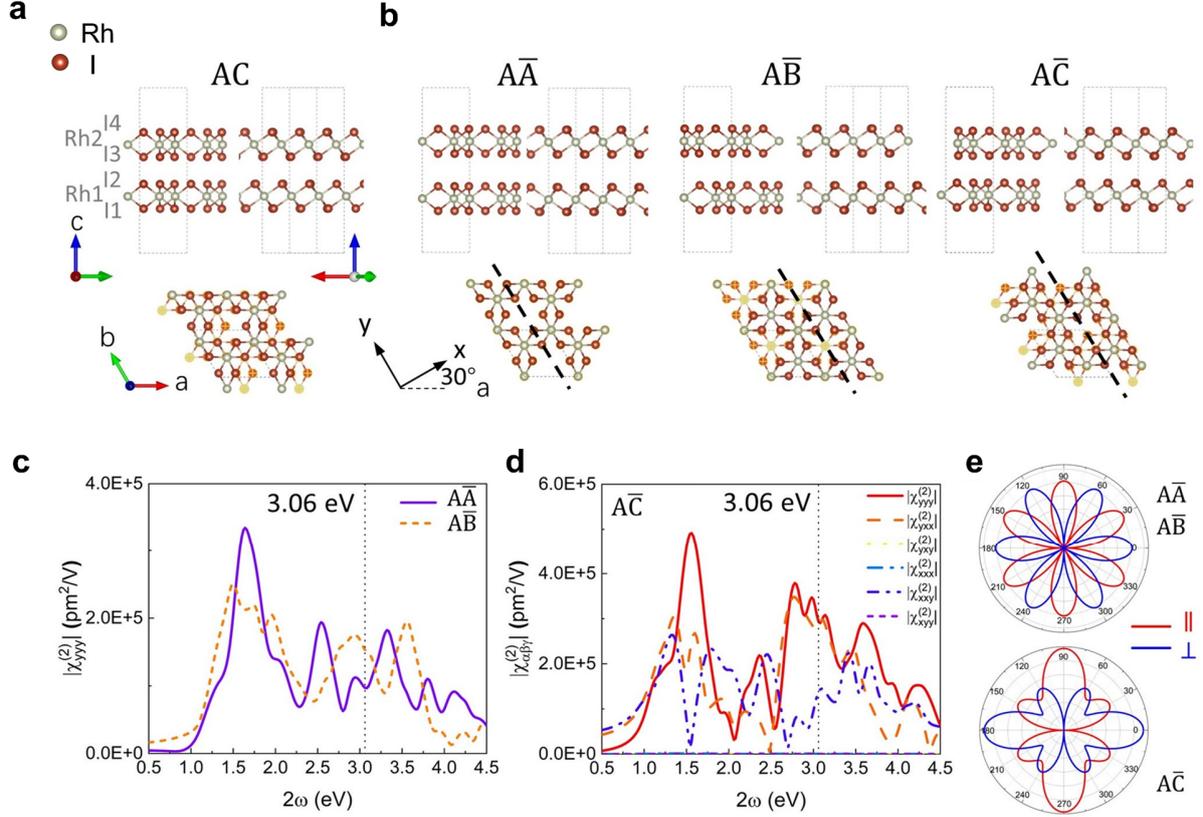

**Fig. 1. Atomic structures and calculated SHG spectra of 2L RhI₃ in different stacking orders.**
**a.** Atomic structure of 2L RhI$_3$ in pristine AC stacking order. The grey dashed lines illustrate the primitive cell. **b.** Atomic structures of 2L RhI$_3$ in A$\bar{\text{A}}$, A$\bar{\text{B}}$, and A$\bar{\text{C}}$ stacking orders. The bar means the top layer is flipped over with respect to AA and AB stacking in **Fig. S1**. Lattice indices and cartesian axes are shown by arrows. The black dashed lines represent the projection of a mirror plane on *xy* plane. Bottom atoms are represented by yellow crosses. **c.** Calculated SHG spectra corresponding to $\left|\chi^{(2)}_{yyy}\right|$ of 2L A$\bar{\text{A}}$ and A$\bar{\text{B}}$ RhI$_3$. **d.** Calculated SHG spectra corresponding to the absolute values of all independent second-order susceptibility elements, $\left|\chi^{(2)}_{yyy}\right|$, $\left|\chi^{(2)}_{yxx}\right|$, $\left|\chi^{(2)}_{yxy}\right|$, $\left|\chi^{(2)}_{xxx}\right|$, $\left|\chi^{(2)}_{xxy}\right|$, and $\left|\chi^{(2)}_{xyy}\right|$ of 2L A$\bar{\text{C}}$ RhI$_3$. The black dashed lines mark $2\omega = 3.06$ eV which is related to the SHG measurement with the excitation laser at 810 nm in later context. **e.** Calculated SHG intensity ($2\omega = 3.06$ eV) plotted as a function of the polarization direction of incident light. The red and blue lines correspond to that the polarizations of incident and outgoing light are parallel and perpendicular, respectively.

Pristine RhI$_3$ is a monoclinic crystal and its lattice parameters are *a*=6.770 Å, *b*=11.720 Å, *c*=6.830 Å, and *β*=109.30°. It is a vdWs material with the stacking order (AC) as shown in **Fig. 1a**. The rhodium atoms in each layer are on top of the nearest iodine sites in the neighboring layer with a tiny in-plane shift. Each layer is composed of edge-sharing interconnected [RhI$_6$] octahedra. The pristine RhI$_3$ single crystal is centrosymmetric (space group: *C2*/m) where the even-order nonlinear optical response is absent, such as SHG. However, it has been reported that stacking faults are quite normal in transition-metal trihalides MX$_3$[23–26] which might break the inversion symmetry. Hence, SHG measurement becomes a powerful method to validate the existence of stacking faults in MX$_3$, and polarized SHG signal is helpful to characterize the types of stacking faults in these systems.

**Table 1. Relative energy, space group, and SHG response of the primitive cells of 2L RhI$_3$ in different stacking orders.**

| Stacking order | Relative energy per atom (meV) | Space group | SHG response |
|:---:|:---:|:---:|:---:|
| AA | 8.3 | P-31m | No |
| AB | 0 | P-3 | No |
| AC | 0.7 | P-1(C2/m)* | No |
| A$\overline{\text{A}}$ | 19.9 | P-62m | Yes |
| A$\overline{\text{B}}$ | 16.2 | P321 | Yes |
| A$\overline{\text{C}}$ | 1.2 | P1(Cm)* | Yes |

*The space groups of the unit cells for AC and A$\overline{\text{C}}$ are shown in the brackets. Both have an extra mirror plane which is not presented in their primitive cell.

Based on density functional theory (DFT) calculations, we simulated various stacking faults *via* bilayer (2L) RhI$_3$ in different stacking orders and calculated their second-order susceptibilities to investigate their nonlinear optical properties. In our work, we only consider common stacking orders generated from in-plane translation and mirror operation (*xy* mirror plane) of one layer, while the stackings containing random twist angles are out of consideration due to their relatively higher energies in general[27]. Beginning from the AC stacking, shifting one RhI$_3$ layer could generate two more general stacking orders: AA and AB. As shown in **Fig. S1**, one layer is a repetition of the other one along the out-of-plane direction in AA stacking, while Rh atoms are on top of the hollow sites of the other layer in AB stacking.

For 2L RhI$_3$, AB stacking is the most stable phase, while AC stacking is 0.7 meV per atom higher in energy than AB stacking, making it secondly stable (**Table 1**). SHG response is forbidden in all three phases because of their centrosymmetric structures (**Table 1**). To break the inversion symmetry, one mirror operation is introduced in one of the RhI$_3$ layers considering the Rh atomic plane as the mirror plane. Three phases in A$\overline{\text{A}}$, A$\overline{\text{B}}$, and A$\overline{\text{C}}$ stacking orders are shown in **Fig. 1b** which correspond to AA, AB, and AC stackings with the top RhI$_3$ layer flipped over, respectively. Among all stackings, 2L A$\overline{\text{C}}$ RhI$_3$ is only 1.2 meV per atom higher in energy than the most stable phase, and hence, the existence of stacking faults related to A$\overline{\text{C}}$ stacking in RhI$_3$ bulk is possible naturally.

Inversion symmetry breaking due to stacking faults can induce SHG response in vdWs material[28]. We predict the SHG susceptibility of 2L RhI$_3$ in different stacking orders *via* the first principle calculations to probe the stacking-fault-induced nonlinear optical properties. In nonlinear optics, the general expression of the optical response can be written as the expansion of polarization $P(t)$ in the power series of electric field $E(t)$[29]:

$$P(t) = \varepsilon_0 [\chi^{(1)} E(t) + \chi^{(2)} E^2(t) + \chi^{(3)} E^3(t) + \cdots] \qquad (1)$$

where $\varepsilon_0$ is absolute permittivity; $\chi^{(1)}$, $\chi^{(2)}$, and $\chi^{(3)}$ are the first-, second-, and third-order optical susceptibility, respectively. SHG is a response that the frequency of outgoing light is double of the incident light, corresponding to the second-order susceptibility which is given below under the independent particle approximation (IPA)[30–32].

$$\chi^{(2)}_{\alpha\beta\gamma}(-2\omega, \omega, \omega) = \frac{-ie^3 \hbar^4}{\widetilde{\omega}^3 m^3 S} \sum_{k, nml} \frac{1}{\omega_{mn,k} - 2\widetilde{\omega}}$$

$$\times \left[ \frac{f_{nl,k}\Lambda^{\alpha}_{nm,k}\{\Lambda^{\beta}_{ml,k}\Lambda^{\gamma}_{ln,k}\}}{\omega_{ln,k} - \widetilde{\omega}} + \frac{f_{ml,k}\Lambda^{\alpha}_{nm,k}\{\Lambda^{\gamma}_{ml,k}\Lambda^{\beta}_{ln,k}\}}{\omega_{ml,k} - \widetilde{\omega}} \right] \quad (2)$$

Here, $\{\Lambda^{\beta}_{ml,k}\Lambda^{\gamma}_{ln,k}\} = \frac{1}{2}[\Lambda^{\beta}_{ml,k}\Lambda^{\gamma}_{ln,k} + \Lambda^{\gamma}_{ml,k}\Lambda^{\beta}_{ln,k}]$, where $\Lambda^{\lambda}_{nm,k} = \langle n\mathbf{k}|\hat{p}^{\lambda}|m\mathbf{k}\rangle$ ($\lambda = x, y,$ or $z$.) are the momentum matrix elements. $\alpha$ is the polarization of incident light, $\beta$ and $\gamma$ are the polarization of outgoing light, which can take $x, y,$ or $z$. $\widetilde{\omega} = \omega + i\eta$ where $\omega$ is the photon energy of incident light and $\eta$ is a small positive value which is taken to be 0.05 eV to describe the adiabatic switching-on of the electromagnetic field[33]. $\omega_{mn,k} = \epsilon_{n,k} - \epsilon_{m,k}$ is the energy difference between the Bloch states $|n\mathbf{k}\rangle$ and $|m\mathbf{k}\rangle$. $f_{nl,k}$ is the difference between the occupation numbers of $|l\mathbf{k}\rangle$ and $|n\mathbf{k}\rangle$. Our numerical results of $\chi^{(2)}_{\alpha\beta\gamma}$ are normalized to the in-plane area $S$ of the 2L RhI$_3$ primitive cell, resulting in a unit of pm$^2$/V. To be clear, it is different from the conventional unit of SHG for bulk (pm/V) which is normalized by the volume.

Calculated SHG spectra are the plots of the magnitudes of second-order optical susceptibilities $|\chi^{(2)}_{\alpha\beta\gamma}(-2\omega, \omega, \omega)|$ against the photon energy of outgoing light $2\omega$. Considering that the polarization of light in out-plane direction of 2D materials is negligible in experiment as well as the intrinsic permutation symmetry of tensor elements, there are six independent elements left: $\alpha\beta\gamma = xxx, xxy, xyy, yxx, yxy,$ and $yyy$. Since the atomic structures of 2L A$\overline{\text{A}}$, A$\overline{\text{B}}$, and A$\overline{\text{C}}$ RhI$_3$ have a mirror plane $yz$ as depicted by the dashed line in their top views (**Fig. 1b**), there are only three non-vanishing independent components $\chi^{(2)}_{yyy}, \chi^{(2)}_{yxx},$ and $\chi^{(2)}_{xxy}$ in total. **Fig. 1c** exhibits the SHG spectra $|\chi^{(2)}_{yyy}|$ for 2L A$\overline{\text{A}}$ and A$\overline{\text{B}}$ RhI$_3$ which is the same as $|\chi^{(2)}_{yxx}|$ and $|\chi^{(2)}_{xxy}|$, because the three-fold rotational symmetry in 2L A$\overline{\text{A}}$ and A$\overline{\text{B}}$ RhI$_3$ results in $\chi^{(2)}_{yyy} = -\chi^{(2)}_{yxx} = -\chi^{(2)}_{xxy}$. From the SHG spectra of 2L A$\overline{\text{C}}$ RhI$_3$ in **Fig. 1d**, we can see that $|\chi^{(2)}_{yyy}|$ of 2L

A$\overline{C}$ RhI$_3$ is slightly larger but in the same order as that of 2L A$\overline{A}$ and A$\overline{B}$ RhI$_3$, and $\left|\chi_{xxx}^{(2)}\right|$, $\left|\chi_{xyy}^{(2)}\right|$, and $\left|\chi_{yxy}^{(2)}\right|$ are zero. All phases show a strong peak at ~1.60 eV, meaning when the wavelength of incident light is 1550 nm (0.80 eV), a strong SHG response at 775 nm (1.60 eV) will be generated. Besides, the remarkable SHG response from 1.60 eV to 3.00 eV indicates a stable conversion from infrared light to visible light, which could be used to achieve infrared-light detection by common high-sensitive visible light photodetectors.

Due to the different symmetries in three stackings, there is a clear distinction between their polarized SHG spectra. The predicted polar plots of the SHG intensity ($2\omega = 3.06$ eV) against the polarization direction of incident light for 2L A$\overline{A}$, A$\overline{B}$, and A$\overline{C}$ RhI$_3$ are shown in **Fig. 1e**. This is related to the SHG measurement with the excitation laser at 810 nm which is one of common lasers in labs and is used in this work as reported in later context. Both parallel ($\overrightarrow{e_i} \parallel \overrightarrow{e_o}$) and perpendicular ($\overrightarrow{e_i} \perp \overrightarrow{e_o}$) configurations are considered, where $\overrightarrow{e_i}$ and $\overrightarrow{e_o}$ refer to the polarization direction of the incident and outgoing light, respectively. Details of the calculation method can be found in **Supplementary Information (SI) Section S1**. The polar plots of all three structures exhibit a six-petal feature. However, six petals in the plot for A$\overline{C}$ RhI$_3$ are unequal while they are equivalent in A$\overline{A}$ and A$\overline{B}$ stackings due to the existence of three-fold rotational symmetry. When a tiny in-plane shift of one layer is applied, the equality of petals can be broken as illustrated in **Fig. S2**. Hence, the unequal petals of A$\overline{C}$ stacking is because 2L A$\overline{C}$ RhI$_3$ can be considered as a relatively stable structure constructed by a significant in-plane translation of one layer from A$\overline{A}$ or A$\overline{B}$ that breaks the rotational symmetry.

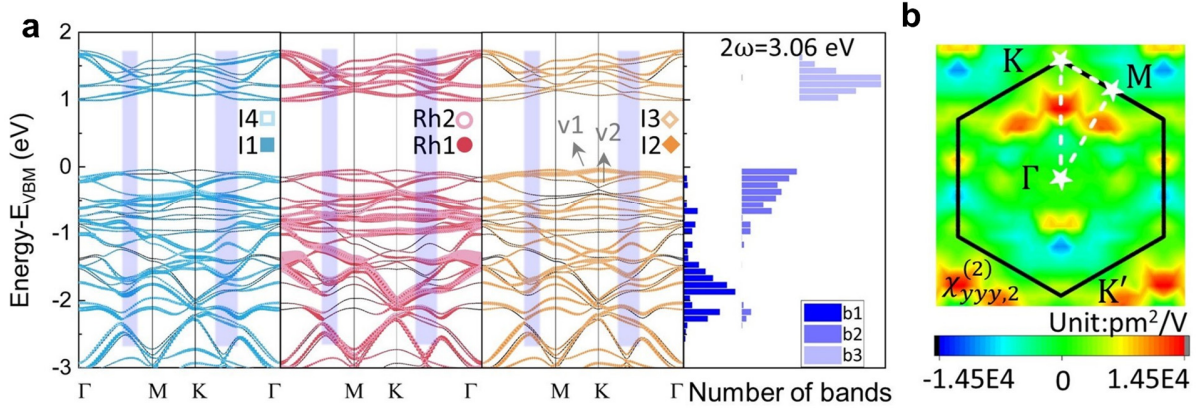

**Fig. 2. Microscopic origin of SHG response of 2L A$\bar{\text{C}}$ RhI$_3$. a.** Band structure with the projections of atomic orbitals and the statistical analysis of the bands corresponding to the transitions contribute dominantly to the SHG susceptibility element $\chi^{(2)}_{yyy}$ (the absolute value larger than 10% of its imaginary part $\chi^{(2)}_{yyy,2}$) at $2\omega = 3.06$ eV. **b.** The *k*-point-resolved imaginary part of the SHG susceptibility element $\chi^{(2)}_{yyy,2}$ at $2\omega = 3.06$ eV. The first Brillouin zone (BZ) and the high symmetry path are shown by black solid lines and white dashed lines. The color bar is at the bottom.

To elucidate the microscopic origin of SHG response in 2L A$\bar{\text{C}}$ RhI$_3$, the summation in **Equation 2** can be splitted by bands and *k* points. The band-resolved SHG can be calculated by

$$\chi^{(2)}_{\alpha\beta\gamma,nml} = \frac{-ie^3\hbar^4}{\widetilde{\omega}^3 m^3 S} \sum_{k,n,m,l\in\{b1,b2,b3\}} \frac{1}{\omega_{mn,k} - 2\widetilde{\omega}}$$

$$\times \left[ \frac{f_{nl,k}\Lambda^{\alpha}_{nm,k}\{\Lambda^{\beta}_{ml,k}\Lambda^{\gamma}_{ln,k}\}}{\omega_{ln,k} - \widetilde{\omega}} + \frac{f_{ml,k}\Lambda^{\alpha}_{nm,k}\{\Lambda^{\gamma}_{ml,k}\Lambda^{\beta}_{ln,k}\}}{\omega_{ml,k} - \widetilde{\omega}} \right] \qquad (3)$$

where the group $\{n, m, l\}$ takes all possible combinations of three random bands $b1$, $b2$, and $b3$. The *k*-resolved SHG is the summation over bands at each *k* point. We focus on two large susceptibility elements, $\chi^{(2)}_{yyy}$ and $\chi^{(2)}_{yxx}$, of 2L A$\bar{\text{C}}$ RhI$_3$ at 3.06 eV. The imaginary part of the susceptibility $\chi^{(2)}_{yyy,2}$ of 2L A$\bar{\text{C}}$ RhI$_3$ at 3.06 eV is the dominant contribution to the absolute value of $\chi^{(2)}_{yyy}$ because it is ~9 times larger than the real part $\chi^{(2)}_{yyy,1}$. Hence, we investigate the bands included in the transitions that play a significant role in the imaginary part. To be clear, the quantities of related combinations $\{b1, b2, b3\}$ of which $\left|\chi^{(2)}_{yyy,nml,2}\right|$ is larger than 10% of

$\left|\chi_{yyy,2}^{(2)}\right|$ are counted in the bar plots (**Fig. 2a**). The width of each bar is 0.1 eV. The statistical analyses demonstrate that the dominant contributions of $\chi_{yyy,2}^{(2)}$ are from three groups of bands at -2.0 to -1.5 eV, -0.5 to 0 eV, and 1.0 to 1.5 eV, respectively, which is close to the resonant condition of SHG in **Equation 3**. The resonant conditions tell that when the electronic transition energies between bands match with the energies of the incident or outgoing light (~1.53 eV or ~3.06 eV), the denominator will be close to zero which could induce a huge resonant enhancement of SHG response. The *k*-point-resolved $\chi_{yyy,2}^{(2)}$ in **Fig. 2b** indicates that the electronic states along the high symmetry path $\Gamma - K$, $\Gamma - K'$, and $\Gamma - M$ covered by the blue rectangles in the band structure contribute prominently to SHG response at 3.06 eV. Furthermore, looking into the orbital projections on the pivotal states (**Fig. 2a**), it is found bands $b1$ and $b3$ mainly belong to Rh layers, bands $b2$ at the band edge primarily originate two neighboring I layers at the interface between two layers (i.e., I2 and I3), and bands $b2$ at ~ -0.50 eV are composed of orbitals in I1, I4, and Rh layers. As revealed in **Fig. S3**, $\chi_{yxx}^{(2)}$ mainly originates from the same band groups as $\chi_{yyy}^{(2)}$ which leads to resonant enhancement at 3.06 eV. However, as the momentum matrix elements included are in different directions, the *k*-resolved $\chi_{yxx}^{(2)}$ show some distinctions that states along $\Gamma - M$ no longer contribute.

**Identifications of stacking faults**

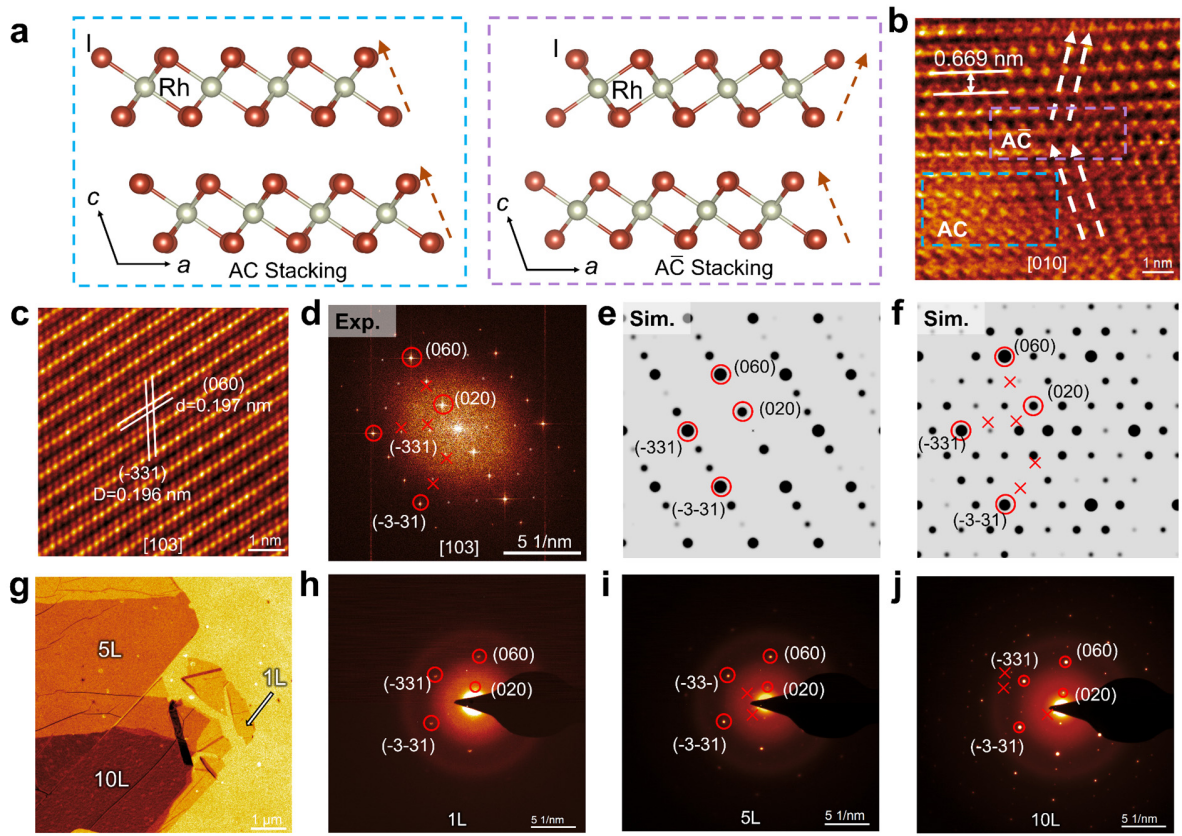

**Fig. 3. Structural analyses of RhI₃ nanoflake. a.** Atomic structures of the pristine phase (AC, dashed blue box) and stacking-fault (AC̄, dashed purple box) phase. The yellow dashed arrows mark the stacking orientation of RhI₃ layers. **b.** Cross-section high-resolution transmission electron microscopy (HRTEM) image of RhI₃ nanoflake viewed along the [010] zone axis. The white dashed arrows mark the stacking orientation of RhI₃ layers, showing a distinct "V" shape. The interlayer spacing (0.669 nm) is labelled. Lattice *a* and *c* axes of bulk RhI₃ are shown. **c.** HRTEM image of RhI₃ nanoflake along the [103] zone axis. Lattice spacing of (060) and (-331) planes are 0.197 and 0.196 nm, respectively. **d.** Corresponding fast Fourier transform (FFT) pattern for HRTEM image in (**c**). **e.** Simulated electron diffraction spots of pristine phase AC̄ RhI₃ along the [103] zone axis. **f.** Simulated electron diffraction spots of 6L RhI₃ with one AC̄ stacking fault along the [103] zone axis in **Fig. S7**. **g.** Low magnification transmission electron microscopy (TEM) image of a RhI₃ nanoflake consisting of regions in different thicknesses. Selected area electron diffraction (SAED) patterns of monolayer (1L) RhI₃ (**h**), 5L RhI₃ (**i**), and 10L RhI₃ (**j**), taken along the [103] zone axis.

Based on the predictions above, stacking faults corresponding to different stacking orders probably exist and might generate significant SHG response. To discover the evidence, we synthesized RhI₃ crystals using a flux method and studied the SHG response of RhI₃ which should be a centrosymmetric system intrinsically. As shown in **Fig. 3**, a comprehensive analysis using transmission electron microscope (TEM) was conducted to explore the stacking characteristics of $RhI_3$ crystal. The low magnification TEM image and corresponding selected area electron diffraction (SAED) patterns for cross-section $RhI_3$ nanoflake (**Fig. S4**), projected along the [010] zone axis, reveal the layered structure of $RhI_3$ crystal. Additionally, the energy-dispersive spectroscopy (EDS) analysis indicates an approximately atomic ratio of 1:3 for Rh to I in the chemical composition (**Fig. S5**), confirming the compositional purity of the $RhI_3$ crystals (see details in **SI Section S2**). Since $A\overline{C}$ stacking has been predicted as the most energetically favorable configuration that breaks the inversion symmetry, we take $A\overline{C}$ stacking as an example to analyze the stacking faults in $RhI_3$. **Fig. 3a** illustrates the atomic structure of pristine AC stacking phase and $A\overline{C}$ stacking fault phase, revealing a distinguish transition in stacking order associated with $A\overline{C}$ stacking fault. Cross-section high-resolution transmission electron microscope (HRTEM) image of the $RhI_3$ nanoflake is shown in **Fig. 3b**, which displays a clear layer arrangement of (001) planes with resolved interlayer spacing of about 0.669 nm. It is worth noting that stacking faults within $RhI_3$ nanoflake corresponding to one layer flipped over are observed directly in the HRTEM image as depicted by the white arrows in **Fig. 3b**. The white arrows illustrate the stacking orientations of layers below and above the region of stacking faults, exhibiting a "V" shape which demonstrates an obvious change in the stacking direction on two sides caused by the stacking faults. Since the layers away from stacking faults prefer to follow AC stacking, the stacking faults related to simple in-plane translations could not change the stacking orientation on two sides. Hence, such dramatic change in stacking direction might originate from the mirror operation (*xy* mirror plane) of one layer at the interface (dashed purple box), while in the regions beyond the interface, layers maintain the stacking order as pristine phase RhI$_3$ but change to another stacking direction.

The presence of stacking faults was further uncovered *via* the analyses of electron diffraction patterns. HRTEM image along [103] zone axis, as depicted in **Fig. 3c and Fig. S6**, does not show Moiré patterns[34,35], suggesting the absence of rotational stacking faults. The stacking faults in RhI$_3$ nanoflake only arise from the mirror operation rather than the rotation operation. The corresponding FFT pattern extracted from the HRETM image presented in **Fig. 3d**, exhibits superlattice characteristics with forbidden spots marked by red crosses, offering compelling evidence of the existence of stacking faults. We take the pristine $AC$ phase and $A\bar{C}$ stacking fault structure as the example to simulate electron diffraction pattern. We have found that the pristine $AC$ phase structure (**Fig. 3e**) lacks the superlattice characteristics, whereas the 6L RhI$_3$ with one $A\bar{C}$ stacking RhI$_3$ structure exhibits the same superlattice characteristics (**Fig. 3f**) as experimental results (see details in **SI Section S3 and Fig. S7**). Such excellent agreement confirms the presence of stacking fault induced by a mirror operation in RhI$_3$ crystal. SAED analysis was performed on RhI$_3$ nanoflake consisting of regions in various thicknesses (**Fig. 3g**), which reveals the pristine $AC$ phase electron diffraction pattern in monolayer (1L) regions (**Fig. 3h**) consistent with the simulated $AC$ phase pattern (**Fig. 3e**). As the thickness increases to 5L and 10 L (**Fig. 3i, j**), the superlattice diffraction spots emerge and become increasingly prominent, further confirming the superlattice diffraction spots resulting from stacking faults.

**Characterization of SHG response**

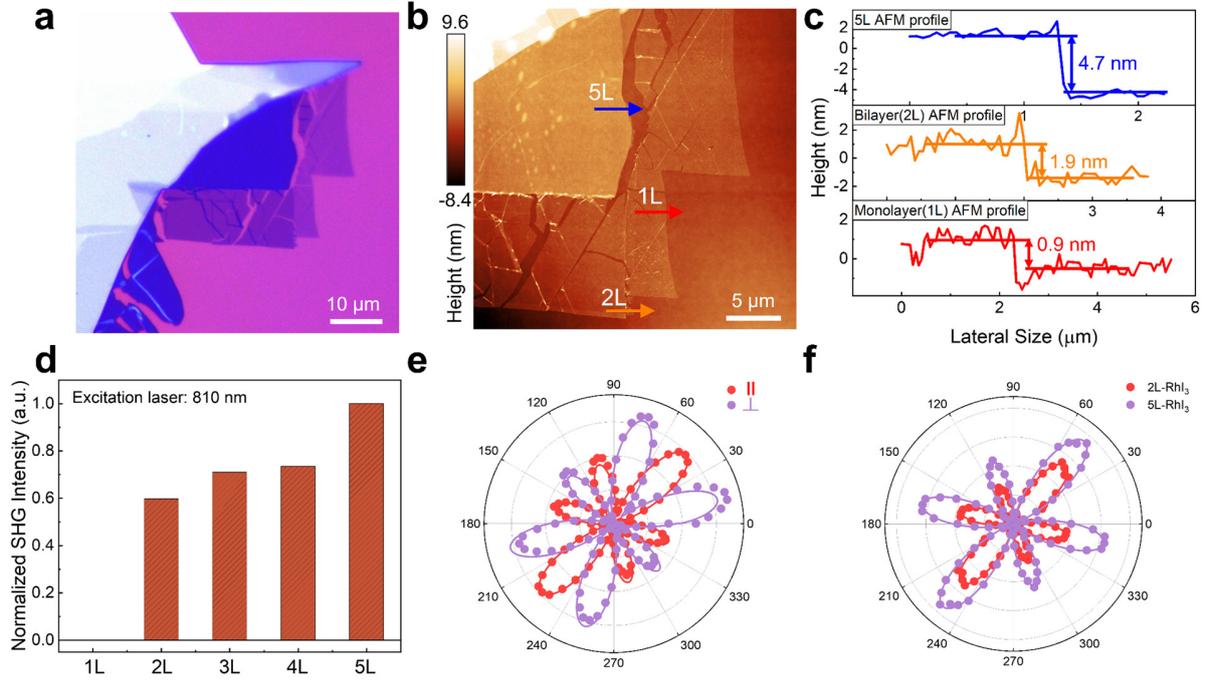

**Fig. 4. SHG measurements of RhI$_3$ nanoflake. a.** Optical microscope (OM) image of a RhI$_3$ nanoflake consisting of regions with different thicknesses. **b.** Corresponding AFM image of the nanoflake in (**a**). **c.** AFM height profiles of 1L, 2L, and 5L RhI$_3$ along the direction depicted by the red, orange, and blue arrows in (**b**), respectively. **d.** SHG intensity of 1L, 2L, 3L, 4L, and 5L RhI$_3$ regions at 810 nm excitation laser. **e.** Polarized SHG spectra of 2L RhI$_3$ flake under parallel (∥) and perpendicular (⊥) configurations. **f.** Polarized SHG spectra of 2L and 5L RhI$_3$ flake under parallel configuration.

For centrosymmetric vdWs materials, SHG measurement is a powerful technique to characterize the regions of stacking faults because the pristine phase has no SHG response. RhI$_3$ nanoflakes are rich in stacking faults which are mainly related to A$\overline{C}$ stacking as proved by above structural characterizations. According to the aforementioned first principle calculations, A$\overline{C}$ stacking RhI$_3$ can generate effective SHG responses. Therefore, we employed a femtosecond laser pulse with an excitation wavelength of 810 nm to investigate the second-order nonlinear optical properties of RhI$_3$ nanoflakes. Few-layer RhI$_3$ nanoflakes were exfoliated from crystals by traditional mechanical exfoliation method and transferred to SiO$_2$/Si substrate assisted by PDMS, as shown in optical microscope (OM) image (**Fig. 4a**). Regions with different contrasts correspond to RhI$_3$ with different thicknesses as indicated in the atomic force microscopy (AFM) image (**Fig. 4b**) and the height profiles at different areas (**Fig. 4c**). The thickness of RhI$_3$ flake was determined accurately by atomic force microscopy (AFM) combined with optical contrast analysis to diminish as much error as possible[36] (see details in **SI Section S4 and Figs. S8-S10**).

The SHG signals of RhI$_3$ in different thicknesses were collected as shown in **Fig. 4d**. Compared with few-layer (2L-5L) RhI$_3$, SHG response of 1L RhI$_3$ is undetectable because of its centrosymmetry structure. SHG intensities from 2L~5L RhI$_3$ can be attributed to the interface inversion symmetry breaking due to the stacking faults which is consistent with our predictions in **Fig. 1**. The differences in SHG intensity between various thicknesses are probably because of the different contents of stacking faults and thickness-dependent bandgap of RhI$_3$[26] which influence the resonant condition of SHG. It is worth noting that the SHG response can only be observed in several 2L RhI$_3$ nanoflakes because most maintain AC stacking as grown. Consequently, the study of SHG properties in this work is based on selected RhI$_3$ samples containing stacking faults and exhibiting SHG signal. The SHG mapping of a RhI$_3$ nanoflake in **Fig. S11** exhibits a relatively uniform SHG intensity over the nanoflake, which can be attributed to the two-dimensional planar defect nature of stacking faults rather than point or line defects. Polarized SHG is sensitive to the symmetry of the system which can be different for distinct stacking faults. The polarized SHG intensity of RhI$_3$ was obtained by adjusting the polarization angle $\theta$ of the excitation laser. The polar plots of 2L RhI$_3$ flake in **Fig. 4e** reveal a distinct unequal six-petal pattern in both parallel and perpendicular configurations, which is consistent with the predicted polarized SHG of A$\bar{\text{C}}$ stacking in **Fig. 1e** and hence further confirm the dominant role of A$\bar{\text{C}}$ stacking in SHG response of RhI$_3$. Subsequently, we performed the polarized SHG measurements on 2L and 5L RhI$_3$ nanoflakes under parallel configuration, which show the same unequal pattern as shown in **Fig. 4f**, indicating the presence of $A\overline{C}$ stacking faults in RhI$_3$ regardless of the thickness.

**Pressure-driven tunability of SHG response**

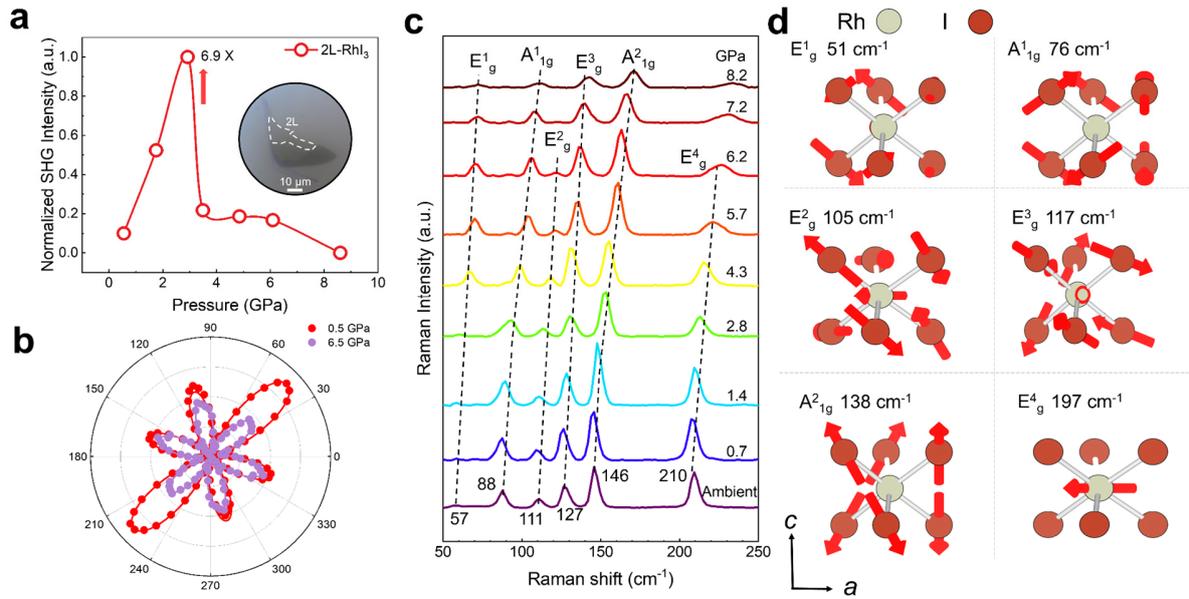

**Fig. 5. SHG and Raman spectra of RhI$_3$ nanoflake under compression. a.** SHG intensity of 2LRhI$_3$ nanoflake as a function of pressure excited by fs laser with a wavelength of 810 nm. The inset shows the optical image of RhI$_3$ nanoflake. **b.** Parallel configuration polarized SHG spectra of 2L RhI$_3$ nanoflake at 0.5 and 6.5 GPa pressure conditions. **c.** Raman spectra of RhI$_3$ nanoflake at different pressure conditions. The dashed lines represent the evolutionary trend of the Raman shift. Peak positions at the ambient condition are given. The irreducible representations of all peaks are labelled. **d.** Vibrational displacements of Raman-active phonon modes for RhI$_3$ are depicted by red arrows.

It has been known that the external pressure can efficiently tune the interlayer interactions in vdWs materials[22] by modulating the interlayer distance. To explore the potential pressure-driven tunability of stacking-fault-induced SHG response in RhI$_3$, we applied hydrostatic pressure on the exfoliated RhI$_3$ nanoflake. As displayed in **Fig. 5a**, the SHG intensity of 2L RhI$_3$ nanoflake increases during compression and results in a 6.9 times enhancement at 2.9 GPa. The pressure-driven SHG enhancement of RhI$_3$ is larger than other vdWs materials listed in **Table S1**. Such SHG enhancement effect under high-pressure conditions was also found in 4L-RhI$_3$ nanoflake at

810 and 1064 nm, where the details can be found in **SI Section S5** and **Figs. S12, S13**. Comparing the parallel configuration polarized SHG at 0.5 and 6.5 GPa (**Fig. 5b**), an unexpected change in the shape of polar plots from an unequal six-petal to an almost equivalent six-petal style is observed. This trend is also found in the polarized SHG measured at 1064 nm when comparing the shapes of plots at lower pressure (**Fig. S14**) and higher pressure (**Fig. S15**). Based on the predicted polarized SHG by DFT calculations, such change might refer to a transition from $A\overline{C}$ to $A\overline{A}$ or $A\overline{B}$ stacking faults. As the interlayer spacing decreases under compression, the enhanced interlayer interaction may cause relative layer sliding to result in a transition in symmetry[37,38]. It is promising to use polarized SHG measurements to detect interlayer sliding and the properties of interfaces in other systems. In addition, the weakened SHG intensity under higher pressure could be attributed to the smaller SHG susceptibility of $A\overline{A}$ and $A\overline{B}$ stackings or the reduction of stacking faults which can break the inversion symmetry in the sample. Another possible reason is that high pressure would significantly modify the band structure[39,40] and hence the resonant condition of SHG susceptibility, which might make the SHG response weaker under the excitation at certain energies.

To examine the evolution of crystal structure in RhI$_3$ under compression, *in-situ* Raman spectra were collected under various pressures. The Raman spectrum of RhI$_3$ nanoflake in ambient pressure exhibits six distinct Raman peaks located at 57 cm$^{-1}$ ($E^1_g$), 88 cm$^{-1}$ ($A^1_{1g}$), 111 cm$^{-1}$ ($E^2_g$), 127 cm$^{-1}$ ($E^3_g$), 146 cm$^{-1}$ ($A^2_{1g}$), and 209 cm$^{-1}$ ($E^4_g$) (**Fig. 5c**). The Raman intensity also exhibits good uniformity in the whole nanoflake area (see details in **SI Section S6** and **Fig. S16**). As the pressure increases, all six Raman peaks shift toward higher wavenumbers (blue shift), accompanied with the peak height reduction and width broadening. The blue shift of Raman frequencies under compression originates from the anharmonic part of the potential energy curve which tells that the bonds stiffen as they shorten upon compression[41]. At lower pressure, the shift in frequencies is tiny meaning that interatomic distance changes little. The external pressure also could significantly reduce the vdWs gap between adjacent atomic layers, consequently enhancing interlayer coupling[42] and changing the phonon frequencies. **Fig. 5d** shows the Raman-active phonon modes for monolayer $RhI_3$ computed from the first principles calculations, illustrating the atomic vibrations related to corresponding Raman peaks. The differences between calculated phonon frequencies and experimental data are within 14 cm$^{-1}$ which are acceptable and might come from the influence of thickness. As shown in **Fig. 5d**, in addition to the in-plane phonon mode $E^4_g$, the remaining five Raman-active modes consist of both in-plane and out-of-plane vibrations of Rh-I bonds, which are more sensitive to the shortened interlayer distance under compression.

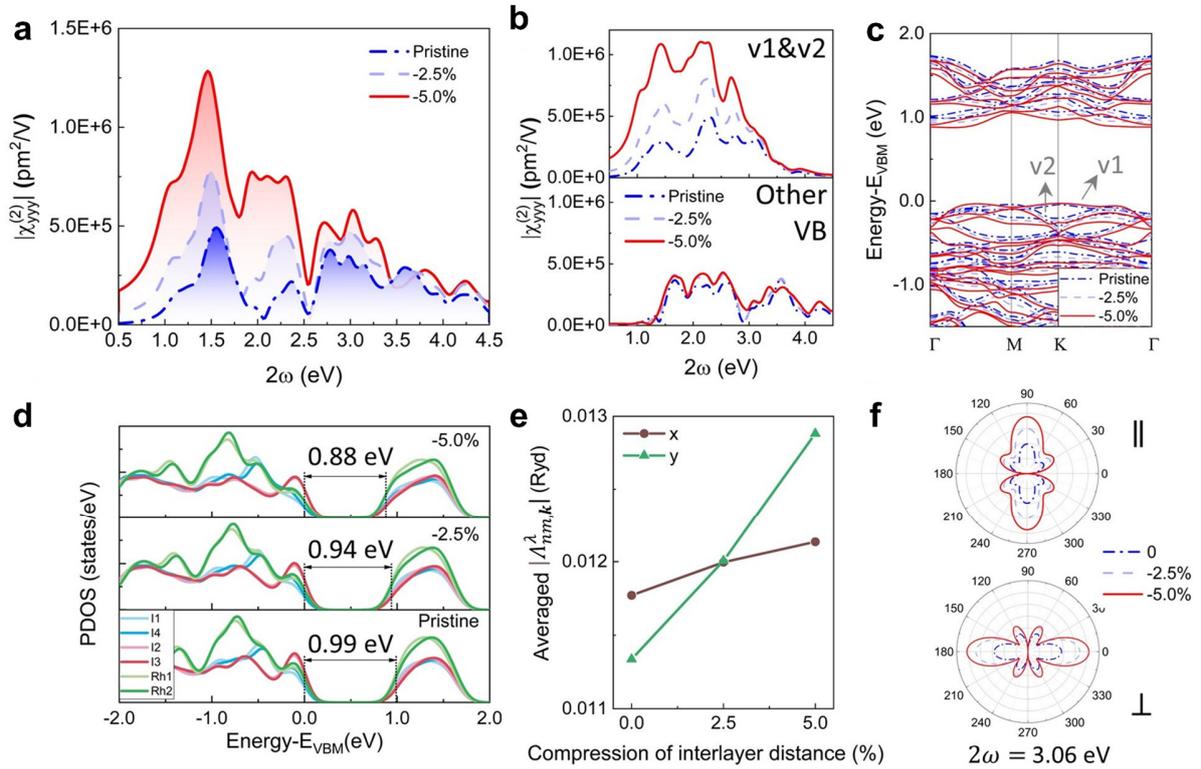

**Fig. 6. Analyses of SHG enhancement effects for 2L A$\bar{C}$ RhI$_3$ under compression. a.** Calculated $|\chi^{(2)}_{yyy}|$ for 2L A$\bar{C}$ RhI$_3$ with and without the compression of interlayer distance. **b.**

Calculated $\left|\chi^{(2)}_{yyy}\right|$ for 2L A$\bar{\text{C}}$ RhI$_3$ under different pressures corresponding to the top two valence bands (v1 and v2), as well as $\left|\chi^{(2)}_{yyy}\right|$ corresponding to other valence bands. **c.** Band structures and **d.** partial density of states (PDOS) of 2L A$\bar{\text{C}}$ RhI$_3$ under different compressions. The bandgap of each system is labeled. **e.** Averaged absolute value of momentum matrix elements $|\Lambda^{\lambda}_{nm,k}|$ ($n \in \{v1, v2\}, m \in \{CB\}, \lambda = x, y$) with and without the compression of interlayer distance. **f.** Calculated SHG intensity ($2\omega = 3.06\text{eV}$) plotted as a function of polarization angle of the incident light for 2L A$\bar{\text{C}}$ RhI$_3$. The polarization of the incident and outgoing light are parallel and perpendicular in the top and bottom polar plots, respectively.

DFT calculations were performed to simulate the evolution of SHG response of RhI$_3$ under different pressures. Because vdWs RhI$_3$ nanoflakes are primarily adsorbed on the surface of diamond anvil during high-pressure experiments, the external pressure is mainly out-of-plane and would result in obvious compression of interlayer spaces[43]. Based on this phenomenon, 2L A$\bar{\text{C}}$ RhI$_3$ with the interlayer distance decreased by 2.5% and 5% are used to simulate the system under low pressure (below 3 GPa)[44]. SHG spectra with and without pressure corresponding to $\left|\chi^{(2)}_{yyy}\right|$, $\left|\chi^{(2)}_{yxx}\right|$, and $\left|\chi^{(2)}_{xxy}\right|$ with and without the compression of interlayer distance are shown in **Fig. 6a**, and **Fig. S17**, respectively. It is found that there is an overall enhancement in SHG intensity when the interlayer distance is shortened, which agrees well with the experimental measurements under pressure below 3 GPa with incident laser at 810 and 1064 nm. Besides an overall enhancement, the SHG peaks in $\left|\chi^{(2)}_{yyy}\right|$ generally shift to lower energies, which demonstrates that the resonant energies of electronic transitions decrease under compression.

To understand the overall enhancement effects and peak shifts, the total $\left|\chi^{(2)}_{yyy}\right|$ is broken into contributions from different states as displayed in **Fig. 6b**. It is found that terms of SHG susceptibility which involve valence bands v1 and v2 play the dominant role in the enhancement of SHG response, while the susceptibility related to other valence states almost remain unchanged under pressure. The summation of the real part between these two susceptibilities equals the total one and so does the imaginary part. Hence, we focus on the electronic transitions related to bands v1 and v2. These two bands are dominant by *p*-orbitals of I2 and I3 layers as shown in the partial density of states (PDOS) in **Fig. S18**. The band structures and partial density of states (PDOS) of 2L A$\bar{C}$ RhI$_3$ in **Fig. 6c, d** indicate that the transition energies between v1 or v2 and the conduction band (CB) are decreased since the CB shifts towards the valence band (VB). This phenomenon corresponds to the narrowed bandgap from 0.99 to 0.94, then to 0.88 eV due to the stronger interlayer interactions as the interlayer spacing shortens. Considering the cubic term of the resonant energy, $\tilde{\omega}^3$, in the denominator of **Equation (2)**, the susceptibility will be significantly magnified when the transition energies shrink. In addition, the magnitudes of momentum matrix elements $\Lambda^\lambda_{nm,k}$ ($n \in \{v1, v2\}, m \in \{CB\}, \lambda = x, y$) become larger as demonstrated by the average values in **Fig. 6e**, which might be because the band broadening under pressure (**Fig. 6d**). Similarly, the absolute value of $\chi^{(2)}_{yxx}$ also exhibit an overall enhancement at different photon energies (**Fig. S17a**). As revealed in **Figs. S19** and **S20**, the enhancement effects below 3.0 eV are mainly from the valence band edge v1 and v2, while the enhanced SHG peak at ~3.4 eV comes from v3 to v7 which are valence bands below v2. The enhancement of the peak at ~3.4 eV can also be attributed to the smaller transition energies as well as the larger momentum matrix elements along *x* direction between v3-7 and CB. To sum up, the significant SHG enhancement effects are the consequence of the narrower bandgap and the broader bandwidth due to stronger interlayer interactions under external pressure. Additionally, the predicted polarized SHG intensities at 3.06 eV (**Fig. 6f**) and 2.33 eV (**Fig. S21**) in parallel and perpendicular configurations maintain the six-petal shape but display enhancement in the magnitudes, which agrees the experimental measurements well (**Fig. S14**).

**Discussion**

In conclusion, we have uncovered the microscopic origins of the stacking-fault-induced SHG response as well as the physical mechanism of its pressure-driven tunability in an intrinsically centrosymmetric vdWs RhI₃ by theoretical and experimental study. Among different stacking orders, $A\overline{C}$ stacking is the most favorable stacking fault which governs the SHG response due to the low formation energy and largest SHG susceptibility as demonstrated by first principle calculations. With the aid of systematical structural characterizations, we have further confirmed the existence of $A\overline{C}$ stacking faults in RhI₃ crystal experimentally. Moreover, the pressure-driven tunability of SHG response is found by applying external pressure using diamond anvil cells and the enhancement factor is up to 6.9. Based on the first principle calculations, such enhancement effects are revealed as the narrowed bandgap and broadened bandwidth due to stronger interlayer interactions. Smaller electronic transition energies and huger momentum matrix elements magnify the SHG susceptibility and hence stronger SHG response. These results provide helpful insights into understanding the relationship between the stacking faults and nonlinear optical effects, as well as its tunability by external stimulation. The prototype demonstrations provided here could serve as a foundation for promoting nonlinear optical efficiency and exploring new avenues in compact optical microelectromechanical systems.

**Methods**

**The first principles calculations**

Density functional theory (DFT) calculations are performed using the generalized gradient approximation (GGA)[45] for the exchange-correlation function as implemented in the plane-wave pseudopotential code, QUANTUM ESPRESSO[46]. The electron-ion interactions are treated by optimized norm-conserving pseudopotentials[47]. The interlayer van der Waals interactions are considered using Grimme's D2 correction[48,49]. A plane-wave kinetic energy cut-off of 70 Ry and a Monkhorst–Pack $k$-point mesh of 6 × 6 × 1 is applied for geometry optimization and self-consistent field calculations, and an energy threshold of $10^{-10}$ Ry is used for the self-consistent cycle. Atomic positions are relaxed until the forces acting on the atoms are less than $10^{-6}$ Ry/Bohr. Periodic boundary conditions with an electrostatic dipole correction are used, with a vacuum length of 16 Å to separate periodic slabs. The partial density of states and SHG susceptibilities are obtained using a denser k-point mesh of $18 \times 18 \times 1$. All valence bands are involved and 203 bands in total are included in calculations to get converged SHG susceptibilities. The phonon modes and frequencies of $RhI_3$ are computed by the density functional perturbation theory (DFPT)[50].

**Materials synthesis and preparation**

$RhI_3$ single crystals were synthesized by a flux method using KI as the flux medium. High-purity Rh powders (99.9%) and I powders (99.9%) were amalgamated with KI and then hermetically sealed within a quartz tube to synthesize the $RhI_3$ single crystals. $RhI_3$ nanoflakes were exfoliated using the conventional mechanical exfoliation method. Subsequently, the exfoliated nanoflakes were transferred to a $SiO_2$/Si substrate or a diamond anvil surface by the PDMS-assisted transfer method. The thickness of $RhI_3$ nanoflake can be determined by optical contrast difference through an optical microscope (Olympus, BX53M) and atomic force microscope (Bruker, Dimension ICON-PT)

**Structural characterization measurements**

The structure analyses of $RhI_3$ nanoflakes were determined by transmission electron microscope (TEM, FEI, Tecnai F30). The focused ion beam (FIB) technique was also employed to prepare the cross-section TEM sample, facilitating an insightful examination of the stacking configuration within the $RhI_3$ flake. The cross-section view unveiled the intricate stacking structure of $RhI_3$.

**High-pressure experiment**

High-pressure characterizations of $RhI_3$ were performed using diamond anvil cell (DAC). Stainless steel gaskets with a diameter of 300 μm and thickness of 250 μm were compressed to 60~80 μm. Subsequently, a laser was employed to create a circular hole with a diameter of 100 μm at the center. Low-fluorescent silicon oil was used as the pressure-transmitting medium. The pressure was determined by the ruby fluorescence method[51]. *In-situ* PL spectra for ruby and *in-situ* Raman spectra for RhI$_3$ were collected by a commercial confocal Raman microscopy (Renishaw, inVia-Reflex) with a 532 nm laser. The *in-situ* SHG spectra at various pressures were measured in a back reflection configuration homemade spectroscopy with a wavelength of 810 nm, and a pulse width of 150 fs laser (or 1064 nm, a pulse width of 80 ps).

**References**

1. Rogée, L. *et al.* Ferroelectricity in untwisted heterobilayers of transition metal dichalcogenides. *Science* **376**, 973–978 (2022).

2. Ma, Q. *et al.* Observation of the nonlinear Hall effect under time-reversal-symmetric conditions. *Nature* **565**, 337–342 (2019).

3. Kang, K., Li, T., Sohn, E., Shan, J. & Mak, K. F. Nonlinear anomalous Hall effect in few-layer WTe$_2$. *Nat. Mater.* **18**, 324–328 (2019).

4. Wu, S. *et al.* Electrical tuning of valley magnetic moment through symmetry control in bilayer MoS$_2$. *Nature Phys* **9**, 149–153 (2013).

5. Bian, G. *et al.* Topological nodal-line fermions in spin-orbit metal PbTaSe$_2$. *Nat Commun* **7**, 10556 (2016).

6. Fu, B.-B. *et al.* Dirac nodal surfaces and nodal lines in ZrSiS. *Science Advances* **5**, eaau6459 (2019).

7. Xu, X. *et al.* Towards compact phase-matched and waveguided nonlinear optics in atomically layered semiconductors. *Nat. Photon.* **16**, 698–706 (2022).

8. Guo, Q. *et al.* Ultrathin quantum light source with van der Waals NbOCl$_2$ crystal. *Nature* **613**, 53–59 (2023).

9. Abdelwahab, I. *et al.* Giant second-harmonic generation in ferroelectric NbOI$_2$. *Nat. Photon.* **16**, 644–650 (2022).

10. Wang, Y. *et al.* Direct electrical modulation of second-order optical susceptibility via phase transitions. *Nat Electron* **4**, 725–730 (2021).

11. Mennel, L. *et al.* Optical imaging of strain in two-dimensional crystals. *Nat Commun* **9**, 516 (2018).

12. Zhang, Y. *et al.* Doping-Induced Second-Harmonic Generation in Centrosymmetric Graphene from Quadrupole Response. *Phys. Rev. Lett.* **122**, 047401 (2019).

13. Karvonen, L. *et al.* Rapid visualization of grain boundaries in monolayer $MoS_2$ by multiphoton microscopy. *Nat Commun* **8**, 1–8 (2017).

14. He, Y.-M. *et al.* Single quantum emitters in monolayer semiconductors. *Nature Nanotech* **10**, 497–502 (2015).

15. Tran, T. T., Bray, K., Ford, M. J., Toth, M. & Aharonovich, I. Quantum emission from hexagonal boron nitride monolayers. *Nature Nanotech* **11**, 37–41 (2016).

16. Cheng, G. *et al.* Emergence of electric-field-tunable interfacial ferromagnetism in 2D antiferromagnet heterostructures. *Nat Commun* **13**, 7348 (2022).

17. Sui, F. *et al.* Sliding ferroelectricity in van der Waals layered γ-InSe semiconductor. *Nat Commun* **14**, 36 (2023).

18. Shan, Y. *et al.* Stacking symmetry governed second harmonic generation in graphene trilayers. *Science Advances* **4**, eaat0074 (2018).

19. Yao, K. *et al.* Enhanced tunable second harmonic generation from twistable interfaces and vertical superlattices in boron nitride homostructures. *Science Advances* **7**, eabe8691 (2021).

20. Sutter, E., Komsa, H.-P., Puretzky, A. A., Unocic, R. R. & Sutter, P. Stacking Fault Induced Symmetry Breaking in van der Waals Nanowires. *ACS Nano* **16**, 21199–21207 (2022).

21. Li, T. *et al.* Pressure-controlled interlayer magnetism in atomically thin $CrI_3$. *Nat. Mater.* **18**, 1303–1308 (2019).

22. Xia, J. *et al.* Strong coupling and pressure engineering in $WSe_2$–$MoSe_2$ heterobilayers. *Nat. Phys.* **17**, 92–98 (2021).

23. Banerjee, A. *et al.* Proximate Kitaev quantum spin liquid behaviour in a honeycomb magnet. *Nature Mater* **15**, 733–740 (2016).

24. Han, X. *et al.* Atomically Unveiling an Atlas of Polytypes in Transition-Metal Trihalides. *J. Am. Chem. Soc.* **145**, 3624–3635 (2023).

25. Sears, J. A. *et al.* Magnetic order in α-$RuCl_3$: A honeycomb-lattice quantum magnet with strong spin-orbit coupling. *Phys. Rev. B* **91**, 144420 (2015).

26. Wang, F. *et al.* Honeycomb $RhI_3$ Flakes with High Environmental Stability for Optoelectronics. *Adv. Mater.* **32**, 2001979 (2020).

27. Wang, Z., Chen, Q. & Wang, J. Electronic Structure of Twisted Bilayers of Graphene/$MoS_2$ and $MoS_2$/$MoS_2$. *J. Phys. Chem. C* **119**, 4752–4758 (2015).

28. Zhang, M. *et al.* Emergent second-harmonic generation in van der Waals heterostructure of bilayer $MoS_2$ and monolayer graphene. *Science Advances* **9**, eadf4571 (2023).

29. Yu, J. *et al.* Giant nonlinear optical activity in two-dimensional palladium diselenide. *Nat Commun* **12**, 1083 (2021).

30. Xuan, F., Lai, M., Wu, Y. & Quek, S. Y. Exciton-enhanced Spontaneous Parametric Down-Conversion in 2D Crystals. *eprint arXiv:2305.08345* (2023).

31. Leitsmann, R., Schmidt, W. G., Hahn, P. H. & Bechstedt, F. Second-harmonic polarizability including electron-hole attraction from band-structure theory. *Phys. Rev. B* **71**, 195209 (2005).

32. Hughes, J. L. P. & Sipe, J. E. Calculation of second-order optical response in semiconductors. *Phys. Rev. B* **53**, 10751–10763 (1996).

33. Rostami, H., Katsnelson, M. I., Vignale, G. & Polini, M. Gauge invariance and Ward identities in nonlinear response theory. *Annals of Physics* **431**, 168523 (2021).

34. Li, Z. *et al.* Direct observation of multiple rotational stacking faults coexisting in freestanding bilayer $MoS_2$. *Sci Rep* **7**, 8323 (2017).

35. Reidy, K. *et al.* Direct imaging and electronic structure modulation of moiré superlattices at the 2D/3D interface. *Nat Commun* **12**, 1290 (2021).

36. Li, H. *et al.* Rapid and Reliable Thickness Identification of Two-Dimensional Nanosheets Using Optical Microscopy. *ACS Nano* **7**, 10344–10353 (2013).

37. Bacaksiz, C. *et al.* Bilayer $SnS_2$: Tunable stacking sequence by charging and loading pressure. *Phys. Rev. B* **93**, 125403 (2016).

38. Tang, L. *et al.* Giant piezoresistivity in a van der Waals material induced by intralayer atomic motions. *Nat Commun* **14**, 1519 (2023).

39. Liu, S. *et al.* Pressure-induced superconductivity and nontrivial band topology in compressed γ-InSe. *Phys. Rev. B* **105**, 214506 (2022).

40. Huang, S. *et al.* Layer-Dependent Pressure Effect on the Electronic Structure of 2D Black Phosphorus. *Phys. Rev. Lett.* **127**, 186401 (2021).

41. Feng, X. *et al.* Giant Tunability of Charge Transport in 2D Inorganic Molecular Crystals by Pressure Engineering. *Angewandte Chemie International Edition* **62**, e202217238 (2023).

42. Xie, X. *et al.* Unveiling layer-dependent interlayer coupling and vibrational properties in $MoTe_2$ under high pressure. *Phys. Rev. B* **108**, 155302 (2023).

43. Zhao, L. *et al.* Probing Anisotropic Deformation and Near-Infrared Emission Tuning in Thin-Layered InSe Crystal under High Pressure. *Nano Lett.* **23**, 3493–3500 (2023).

44. Fang, Y. *et al.* Pressure engineering of van der Waals compound $RhI_3$: bandgap narrowing, metallization, and remarkable enhancement of photoelectric activity. *Materials Today Physics* **34**, 101083 (2023).

45. Perdew, J. P., Burke, K. & Ernzerhof, M. Generalized Gradient Approximation Made Simple. *Phys. Rev. Lett.* **77**, 3865–3868 (1996).

46. Giannozzi, P. *et al.* QUANTUM ESPRESSO: a modular and open-source software project for quantum simulations of materials. *J. Phys.: Condens. Matter* **21**, 395502 (2009).

47. Hamann, D. R. Optimized norm-conserving Vanderbilt pseudopotentials. *Phys. Rev. B* **88**, 085117 (2013).

48. Grimme, S. Semiempirical GGA-type density functional constructed with a long-range dispersion correction. *Journal of Computational Chemistry* **27**, 1787–1799 (2006).

49. Barone, V. *et al.* Role and effective treatment of dispersive forces in materials: Polyethylene and graphite crystals as test cases. *Journal of Computational Chemistry* **30**, 934–939 (2009).

50. Gonze, X. & Lee, C. Dynamical matrices, Born effective charges, dielectric permittivity tensors, and interatomic force constants from density-functional perturbation theory. *Phys. Rev. B* **55**, 10355–10368 (1997).

51. Shen, G. *et al.* Toward an international practical pressure scale: A proposal for an IPPS ruby gauge (IPPS-Ruby2020). *High Pressure Research* **40**, 299–314 (2020).

**Acknowledgments**

This work is supported by the National Key Research and Development Program of China (2019YFA0705201), the National Natural Science Foundation of China (No. 52272146 and No. 52272158), the National Natural Science Foundation of Heilongjiang (No. YQ2021E019), the

**Author contributions**

Conceptualization: W.H, Y.L, C.Y.X

Data curation: Y.L, W.H, B.Z.W

Formal analysis: W.H, F.Y.X

Investigation: Y.L, W.H, B.Z.W, Z.B.Z, J.R.F, J.P.W

Supervision: Y.L, C.Y.X

Writing-original draft: Y.L, H.W

Writing-review & editing: Y.Q.F, L.Z.P, J.Z.W, M.G.Y, F.Q.H, L.Z, Y.L, C.Y.X

**Additional information**

Source data are all provided with this paper. All data are presented in the manuscript or Supplementary information. Correspondence and reasonable requests for materials should be addressed to the corresponding author.

**Competing interests**

All authors declare no competing financial interests.

# Supplementary Information for

# Stacking faults enabled second harmonic generation in centrosymmetric van der Waals RhI$_3$

Yue Liu, Wen He*, Bing-Ze Wu, Feng-Yuan Xuan, Yu-Qiang Fang*, Zheng-Bo Zhong, Jie-Rui Fu, Jia-Peng Wang, Zhi-Peng Li, Jin-Zhong Wang, Ming-Guang Yao, Fu-Qiang Huang, Liang Zhen, Yang Li*, Cheng-Yan Xu*

*Corresponding author. Email: hewenmse@hit.edu.cn (W.H.); fangyuqiang@mail.sic.ac.cn (Y.Q.F); liyang2018@hit.edu.cn (Y.L.); cy_xu@hit.edu.cn (C.Y.X.)

**This file includes:**

Section S1 to S6
Figs. S1 to S21
Table S1
References

## Section S1. Polar plots of SHG intensity

The SHG susceptibility tensor is a rank 3 matrix with 27 elements which can be reduced to 18 elements by applying the intrinsic permutation symmetry, $\chi^{(2)}_{\alpha\beta\gamma} = \chi^{(2)}_{\alpha\gamma\beta}$. In the experiment, the out-of-plane direction of the few-layer RhI$_3$ is along $z$ direction and the dispersion of the linearly polarized incident light is negligible. Hence, the electric field $E_z$ is minimal, and the polarization can be computed as below:

$$P^{(2)} = \varepsilon_0 \begin{pmatrix} \chi^{(2)}_{xxx} & \chi^{(2)}_{xyy} & \chi^{(2)}_{xzz} & \chi^{(2)}_{xyz} & \chi^{(2)}_{xxz} & \chi^{(2)}_{xxy} \\ \chi^{(2)}_{yxx} & \chi^{(2)}_{yyy} & \chi^{(2)}_{yzz} & \chi^{(2)}_{yyz} & \chi^{(2)}_{yxz} & \chi^{(2)}_{yxy} \\ \chi^{(2)}_{zxx} & \chi^{(2)}_{zyy} & \chi^{(2)}_{yzz} & \chi^{(2)}_{zyz} & \chi^{(2)}_{zxz} & \chi^{(2)}_{zxy} \end{pmatrix} \begin{pmatrix} E^2\cos^2(\theta) \\ E^2\sin^2(\theta) \\ 0 \\ 0 \\ 0 \\ 2E\cos(\theta)\sin(\theta) \end{pmatrix} \quad (S1)$$

where $\theta$ is the angle of the polarization direction of the incident light with respect to the crystal $x$-axis. The intensity of SHG polarized in $x$ and $y$ directions is given by

$$I_x \propto \left[\chi^{(2)}_{xxx}\cos^2(\theta) + \chi^{(2)}_{xyy}\sin^2(\theta) + 2\chi^{(2)}_{xxy}\cos(\theta)\sin(\theta)\right]^2 \quad (S2)$$

$$I_y \propto \left[\chi^{(2)}_{yxx}\cos^2(\theta) + \chi^{(2)}_{yyy}\sin^2(\theta) + 2\chi^{(2)}_{yxy}\cos(\theta)\sin(\theta)\right]^2 \quad (S3)$$

while the contribution from $I_z$ is negligible. The SHG intensities with the polarization direction of the incident light parallel and perpendicular to that of the outgoing light are calculated by

$$I_\parallel \propto (\cos(\theta) \quad \sin(\theta)) \begin{pmatrix} I_x \\ I_y \end{pmatrix} \quad (S4)$$

and

$$I_\perp \propto (\sin(\theta) \quad -\cos(\theta)) \begin{pmatrix} I_x \\ I_y \end{pmatrix} \quad (S5)$$

which are plotted in the polar plots of SHG intensity.

## Section S2. Structural characterization

The stacking order of RhI$_3$ nanoflakes was determined by Transmission Electron Microscope (Philips-FEI, Tecnai F30). Then the Focused ion Beam (FIB) technique was employed to prepare the Transmission Electron Microscopy (TEM) sample, facilitating an insightful examination of the stacking configuration within the RhI$_3$ flake. The cross-section view unveiled the intricate stacking structure of RhI$_3$.

The results of the Energy Dispersive Spectroscopy (EDS) elemental scan imaging of the cross-section RhI$_3$ nanoflake prepared by FIB are shown in **Fig. S5**. The high-angle annular dark-field (HAADF) image, reveals the structure composition from bottom to top: SiO$_2$/Si, RhI$_3$, and C film, as shown in **Fig. S5a**. In this region, EDS elemental scan imaging was conducted, yielding the results presented in **Fig. S5b-e**, effectively illustrating the spatial distribution of C, Si, Rh, and I elements. The EDS distributions of Rh and I elements coincide with the RhI$_3$ sample region observed in the HAADF image (**Fig. S5a**). Meanwhile, the elemental composition curve of the scanned region (**Fig. S5f**), from which the Rh to I element ratio is about 1:3, confirms the chemical formula of the experimental material as RhI$_3$.

## Section S3. Electron diffraction simulation

The electron diffraction patterns for the pristine AC and A$\overline{\text{C}}$ stacking fault phase were simulated using the crystallographic software ReciPro[1]. The atomic structures of both pristine bulk and 6L RhI$_3$ with one A$\overline{\text{C}}$ stacking fault were used to simulate the electron diffraction patterns respectively. **Fig. S7** displays the atomic structure of 6L RhI$_3$ with one A$\overline{\text{C}}$ stacking fault. Same simulation parameters were maintained for both phases, including a 300 kV electron wave source that was set to match with the experimental. Additionally, the optics, intensity calculation, and appearance settings were configured as parallel, Kinematical & excitation err, and Gaussian, respectively.

## SectionS4. Identifying the thickness of layered RhI$_3$ flakes

In this work, it is important to accurately determine the thickness of a large number of mechanically exfoliated RhI$_3$ flake samples to explore the stacking faults effect on SHG response. Consequently, atomic force microscopy (AFM) combined with optical contrast analysis was performed to diminish as much error as possible[2,3].

We mainly focused on RhI$_3$ flake with thickness of 1-5L. The representative AFM images as shown in **Fig. S8** exhibit the thickness of different thicknesses. The contrast difference between RhI$_3$ flakes and Si/SiO$_2$ substrate can be obtained simply from the luminance distribution of its color image or the grayscale image of the R, G, or B channels, as shown in **Fig. S9**. Subsequently, the correlation between the thickness and optical contrast was established by:

$$C_D = C - C_S \tag{S6}$$

$$C_{DR} = C_R - C_{SR} \tag{S7}$$

$$C_{DG} = C_G - C_{SG} \tag{S8}$$

$$C_{DB} = C_B - C_{SB} \tag{S9}$$

where $C_D$ is the contrast difference obtained by subtracting $C$ (optical contrast of flake) and $C_S$ (optical contrast of substrate) (**Equation S6**)[4]. $C_{DR}$, $C_{DG}$, and $C_{DB}$ is the contrast difference between the nanosheet and substrate from the R, G, and B channels of grayscale images, respectively. $C_R$, $C_G$, and $C_B$ is the contrast difference of flake from the R, G, and B channels of grayscale images, respectively. $C_{SR}$, $C_{SG}$, and $C_{SB}$ is the contrast difference of substrate from the R, G, and B channels of grayscale images, respectively. Similarly, for the grayscale image (from the R, G, and B channels), the contrast difference between the nanosheet and substrate can also be calculated by subtracting the contrast of the flake from that of the substrate (**Equations S7-S9**).

The grayscale images of the R, G, and B channels of 1L-5L RhI$_3$ flakes were extracted in **Fig. S9**. The optical images of all RhI$_3$ flakes were collected in the same experiment conditions, such as exposure time and white balance parameters. It is worth noting that R and G channels exhibit significant contrast differences in different thickness areas, while B channel is ambiguous, so we only consider the $C_D$, $C_{DR}$, and $C_{DG}$ of RhI$_3$ flakes. Then the correlation profile between the thickness and contrast difference was plotted in **Fig. S10** to accurately determine the thickness of RhI$_3$ flake.

### Section S5. SHG intensity evolution under compression at 1064 nm

We also collected the SHG intensity at various pressure conditions at 1064 nm to explore the SHG intensity evolution under compression (**Fig. S13a**). As shown in **Fig. S13b**, below the pressure condition of 3.72 GPa, the SHG intensity increases gradually with the increase of pressure and reaches the maximum value at 3.72 GPa, which is 3.63 times the ambient condition. As the pressure increases further, the SHG intensity is gradually weakened. Consistent with the results of the *in-situ* SHG test at an excitation wavelength of 810 nm, we also observed a significant SHG enhancement at 1064 nm, which fully demonstrates that the modulation of the structure of RhI$_3$ by high pressure can effectively realize the enhancement of SHG.

### Section S6. Raman mapping measurement

We conducted the Raman mapping measurement on the RhI$_3$ nanoflake to investigate the uniformity of Raman intensity, as shown in **Fig. S16**. It is observed that the intensity mapping outlines of six Raman peaks closely match the contour of the RhI$_3$ nanoflake, indicating uniform intensity distribution. Therefore, the slight displacement variation of incident light on the sample surface during *in-situ* high-pressure tests does not lead to significant differences in Raman signal intensity.

# Supplementary Figures

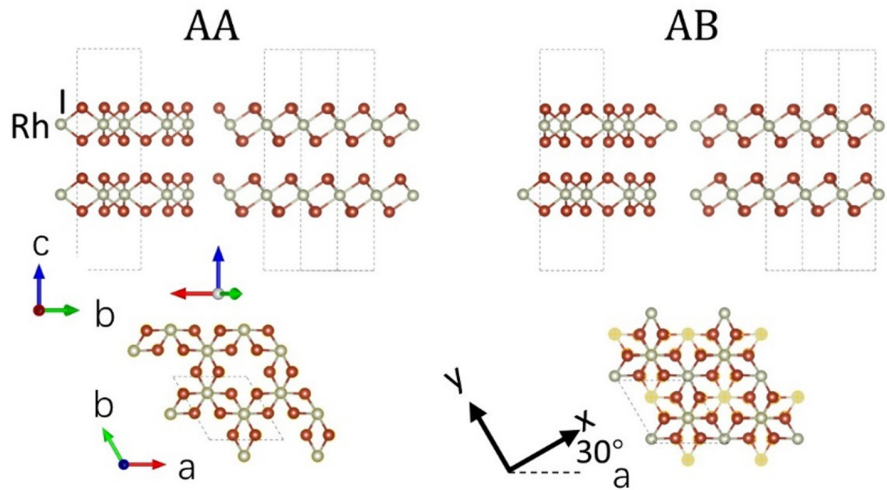

**Fig. S1. Atomic structure of 2L RhI$_3$ with AA and AB stacking.** Lattice indices and cartesian axes are shown by arrows.

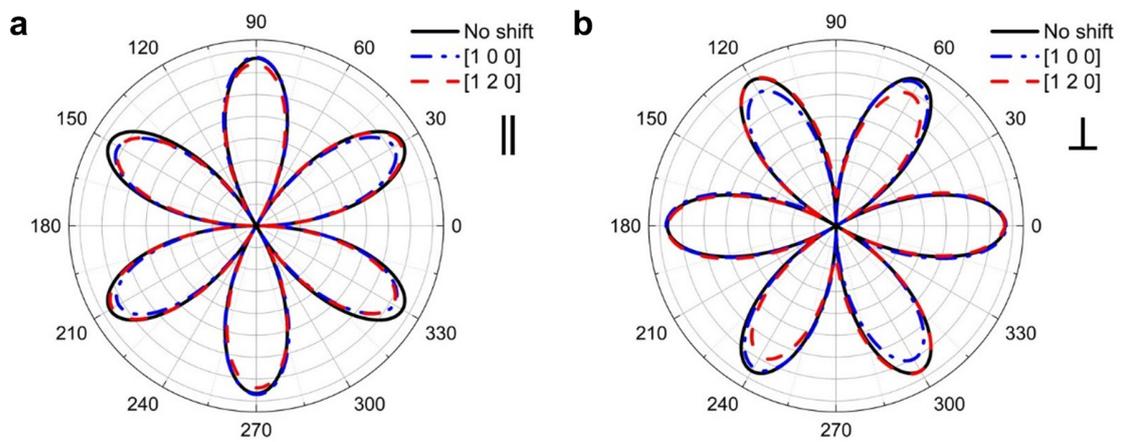

**Fig. S2. Calculated polar plots of SHG intensity of 2L A$\overline{\text{B}}$ RhI$_3$.** The polar plots without and with a 0.2 Å in-plane shift in one layer along [100] and [120] direction. **a.** Parallel configuration ($\vec{e_i} \parallel \vec{e_o}$). **b.** Perpendicular configuration ($\vec{e_i} \perp \vec{e_o}$).

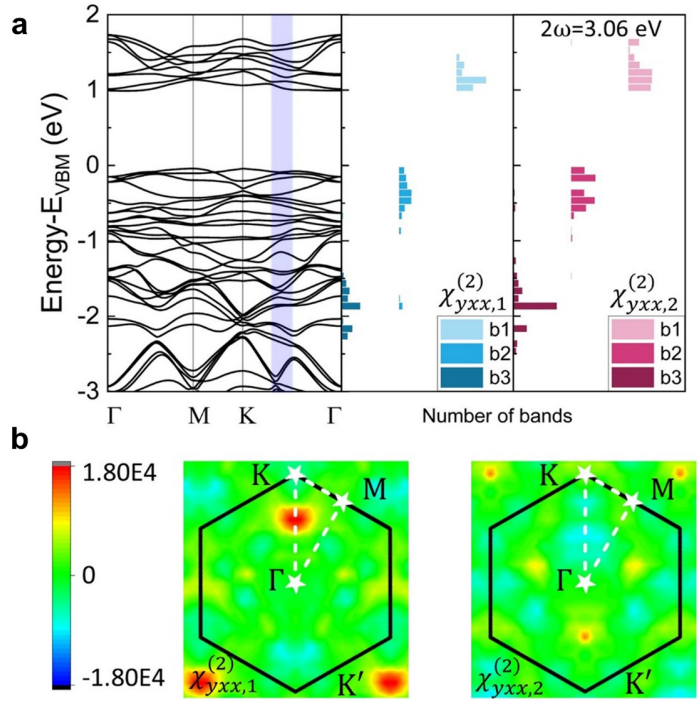

**Fig. S3. Microscopic origin of SHG response of 2L AC̄ RhI₃. a.** Band structure the statistical analysis of the bands corresponding to the transitions contribute dominantly to the real and imaginary part of SHG susceptibility element $\chi^{(2)}_{yxx}$ at 3.06 eV. **b.** The *k*-point-resolved real and imaginary part of SHG susceptibility element $\chi^{(2)}_{yyy}$ at 3.06 eV. The first Brillouin zone (BZ) and the high symmetry path are marked. The color bar is at the bottom.

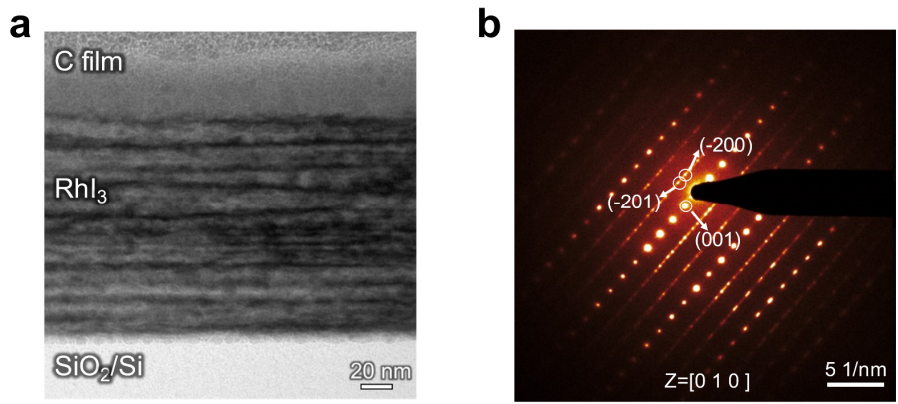

**Fig. S4. Structure characterization. a.** Low magnification transmission electron microscopy (TEM) image of cross-section RhI₃ nanoflake. **b.** Selected area electron diffraction (SAED) patterns along the [010] zone axis.

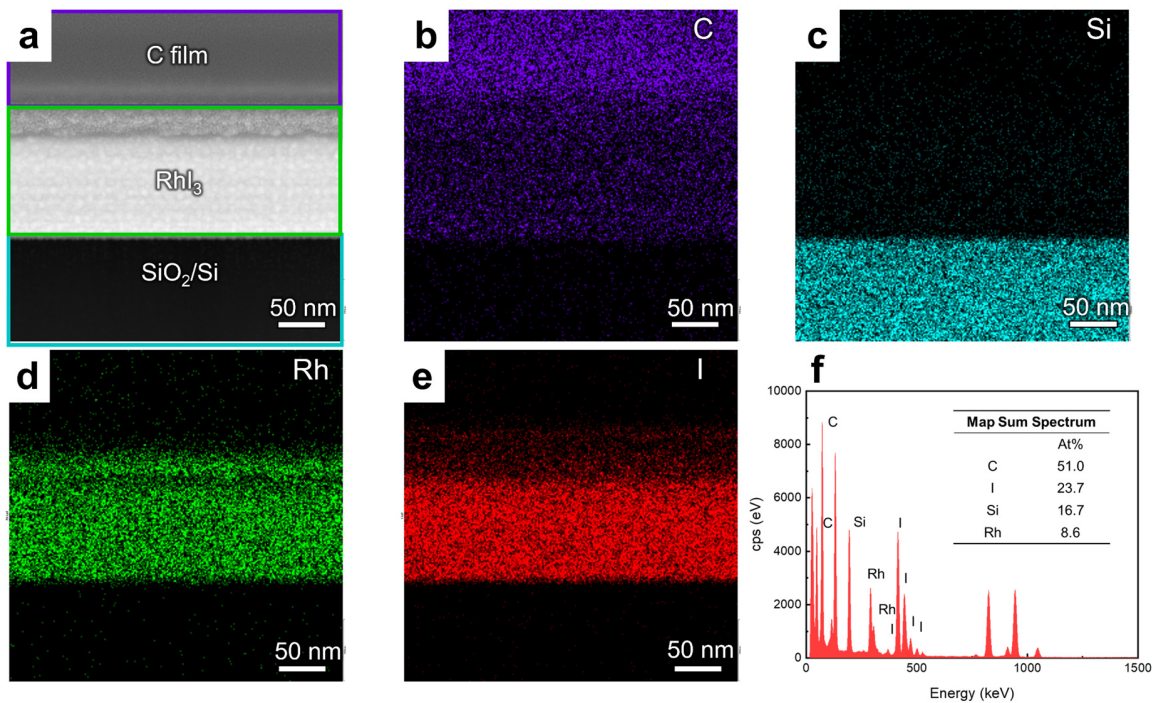

**Fig. S5. HAADF image of cross-sectional RhI₃ flake and corresponding elemental mapping images. a.** HAADF image of the cross-section of RhI$_3$ nanoflake. Corresponding elemental mapping images of C (**b**), Si (**c**), Rh (**d**), and I (**e**). **f.** The profile of element content. It suggests the atomic ratio of Rh: I chemical composition is approximately 1:3, confirming the composition purity of the RhI$_3$ nanoflakes.

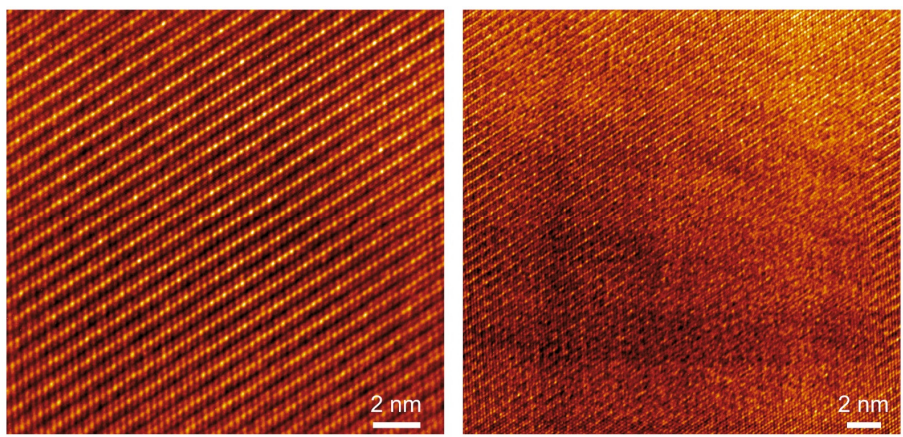

**Fig. S6. Low magnification high-resolution transmission electron microscopy (HRTEM) image of RhI$_3$ nanoflake.**

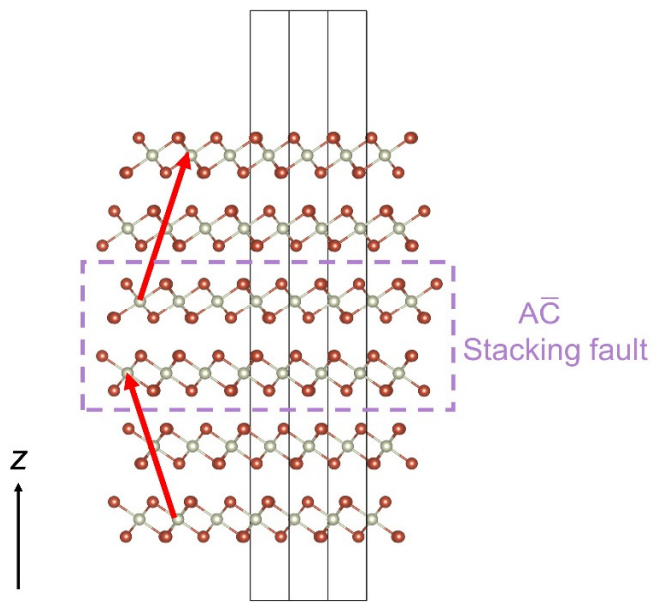

**Fig. S7. Atomic structure of 6L RhI$_3$ with one A$\overline{\text{C}}$ stacking fault**. Black narrow shows the *z*-axis.

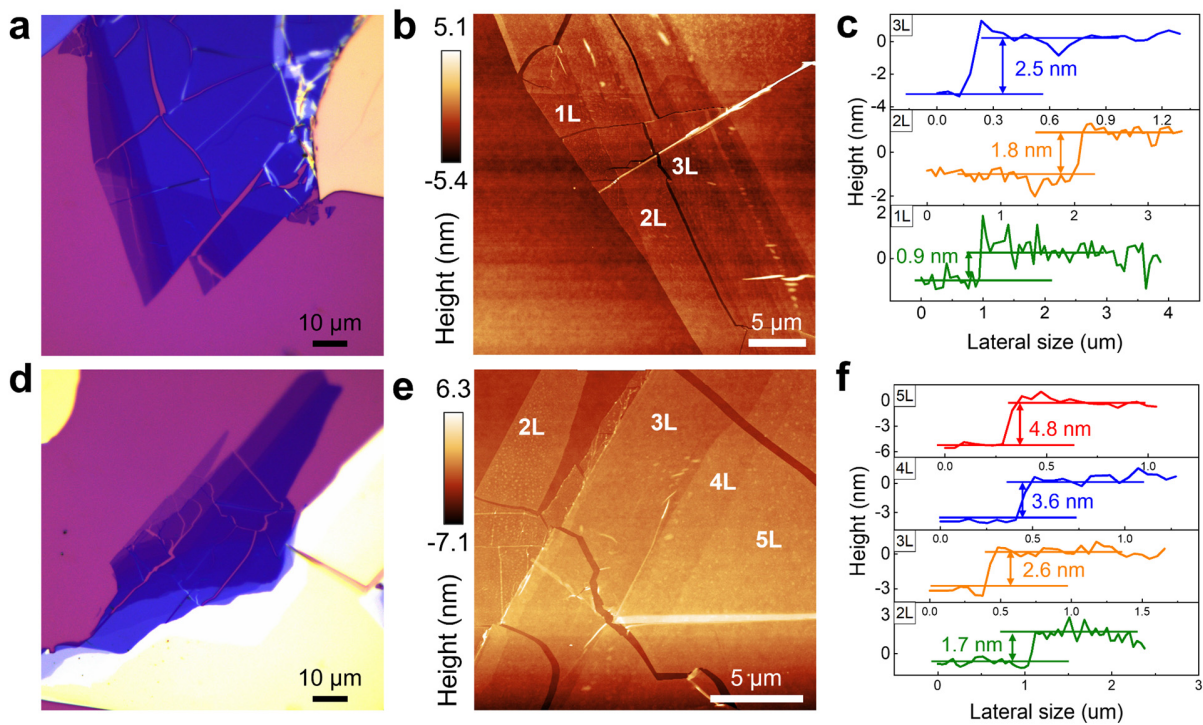

**Fig. S8. Thickness identification by AFM images. a.** Optical image of 1L, 2L, and 3L RhI$_3$ nanoflake. **b.** Corresponding AFM image. **c.** AFM height profiles of 1L, 2L, and 3L RhI$_3$ nanoflakes. **d.** Optical image of 2L, 3L, 4L, and 5L RhI$_3$ nanoflakes. **e.** Corresponding AFM image. **f.** AFM profiles of 2L, 3L, 4L, and 5L RhI$_3$ nanoflakes.

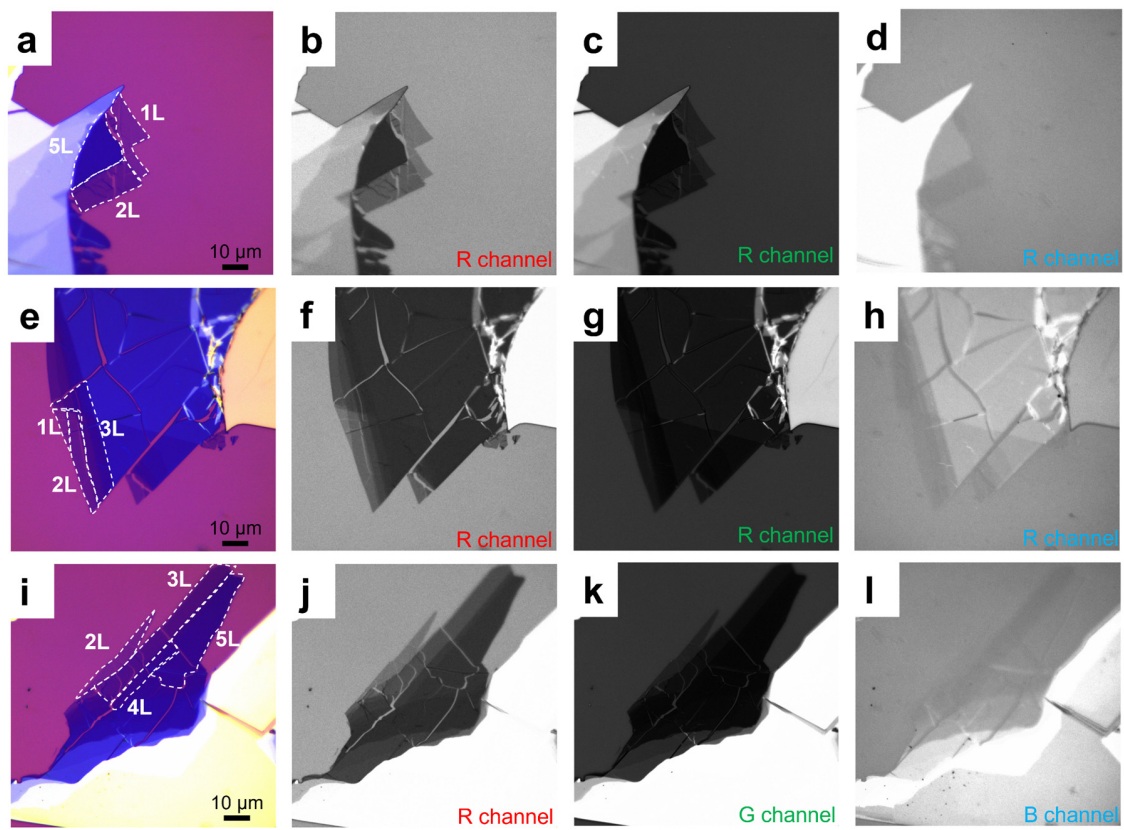

**Fig. S9. Thickness identification by optical contrast difference.** Color (**a**) and grayscale optical images of the (**b**) R, (**c**) G, and (**d**) B channels of 1L, 2L, and 5L RhI$_3$ nanoflakes on SiO$_2$/Si substrate. Color (**e**) and grayscale optical images of the (**f**) R, (**g**) G, and (**h**) B channels of 1L, 2L, and 3L RhI$_3$ flake on SiO$_2$/Si substrate. Color (**i**) and grayscale optical images of the (**j**) R, (**k**) G, and (**l**) B channels of 1L, 2L, and 3L RhI$_3$ nanoflakes on SiO$_2$/Si substrate.

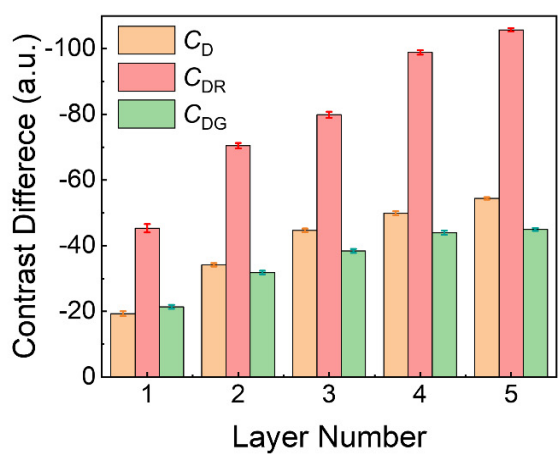

**Fig. S10.** Plot of $C_D$, $C_{DR}$, and $C_{DG}$ value of 1~5L RhI$_3$ nanoflakes on SiO$_2$/Si substrate.

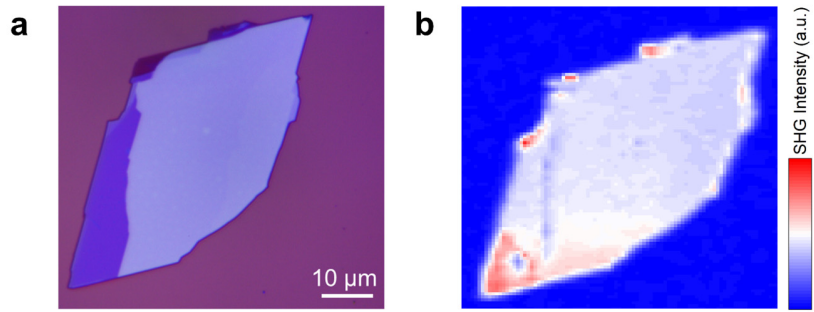

**Fig. S11. SHG spatial mapping measurement. a.** Bright-field optical image. **b.** SHG intensity mapping of RhI$_3$ nanoflake.

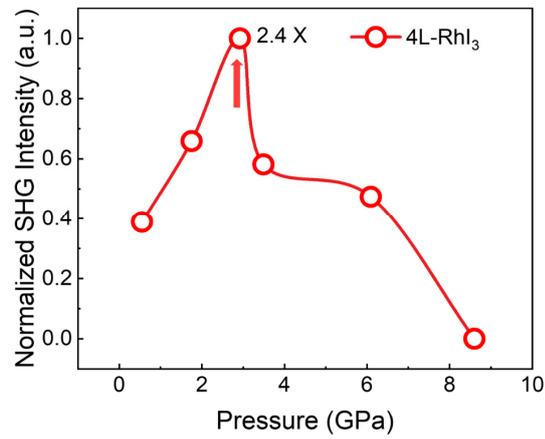

**Fig. S12**. SHG intensity evolution of 4L-RhI$_3$ under compression at 810 nm.

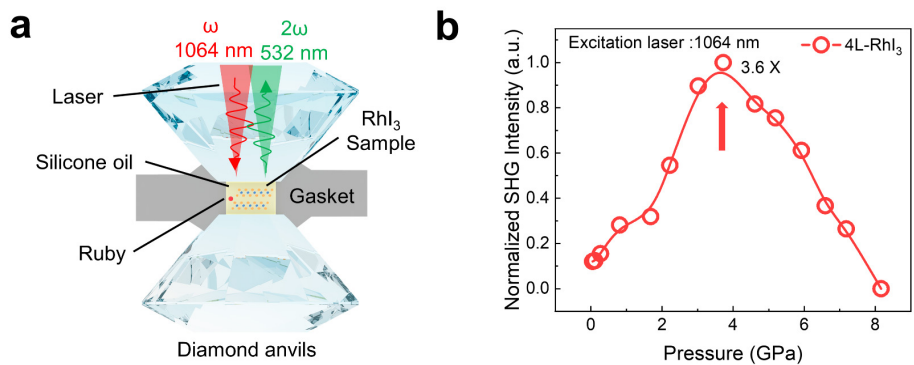

**Fig. S13.** SHG intensity evolution of 4L-RhI$_3$ under compression at 1064 nm. **a.** Schematic of the setup for *in situ* high-pressure SHG measurement. **b.** SHG intensity of a 4L RhI$_3$ nanoflake as a function of pressure.

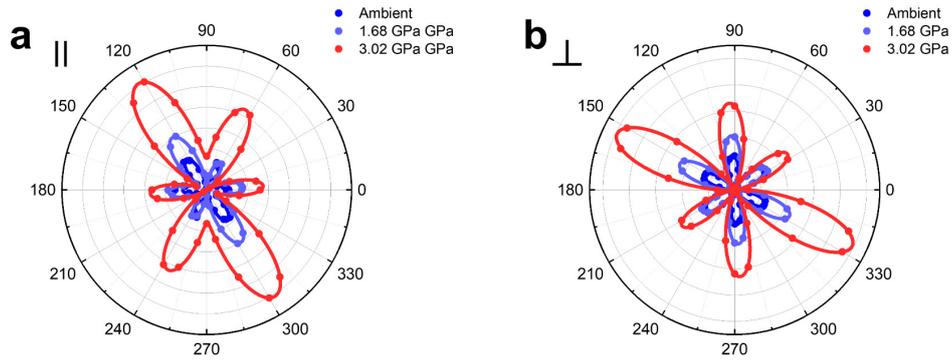

**Fig. S14. Polarized SHG spectra at 1064 nm under lower pressure. a.** Parallel configuration. **b.** Perpendicular configuration.

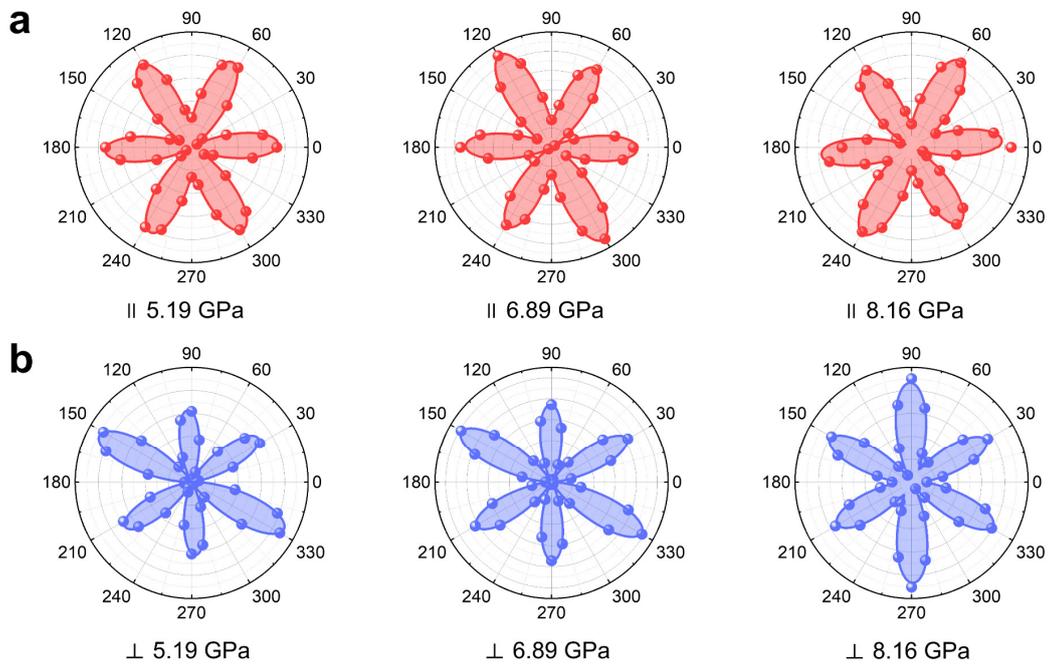

**Fig. S15. Polarized SHG spectra at 1064 nm under higher pressure. a.** Parallel configuration. **b.** Perpendicular configuration.

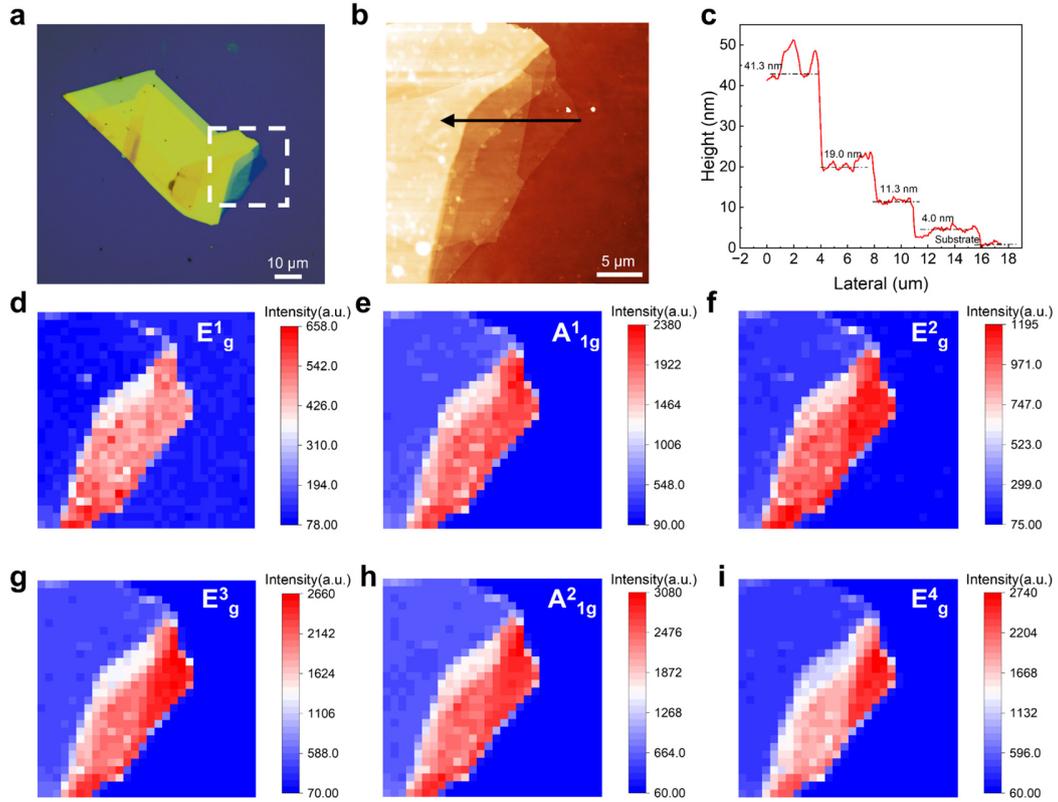

**Fig. S16. Raman intensity mapping of a RhI₃ nanoflake. a.** Optical image of a RhI₃ nanoflake. **b.** AFM image of the area in the white box in (**a**). **c.** The height profile along the black arrow shows the thickness. Raman intensity mapping of phonon mode (**d**) $E^1_g$, (**e**) $A^1_{1g}$, (**f**) $E^2_g$, (**g**) $E^3_g$, (**h**) $A^1_{2g}$, and (**i**) $E^4_g$.

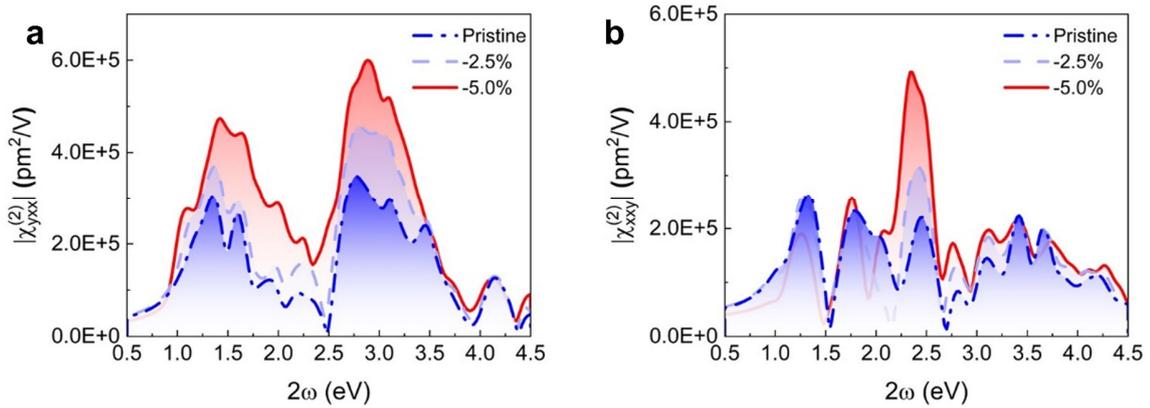

**Fig. S17. Calculated SHG spectra for 2L A$\bar{\text{C}}$ RhI₃ under pressure.** Calculated (**a**) $\left|\chi^{(2)}_{yxx}\right|$ and (**b**) $\left|\chi^{(2)}_{xxy}\right|$ for 2L A$\bar{\text{C}}$ RhI₃ in the pristine structure and with the interlayer distance decreased by 2.5% and 5%.

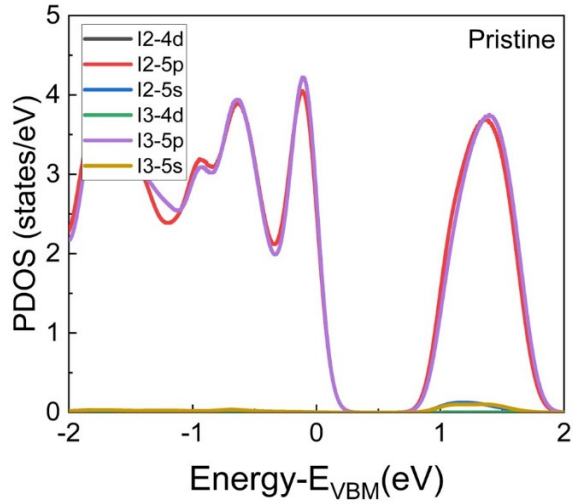

**Fig. S18. Partial density of states (PDOS) of I2 and I3 atomic layers of pristine 2L RhI$_3$.**

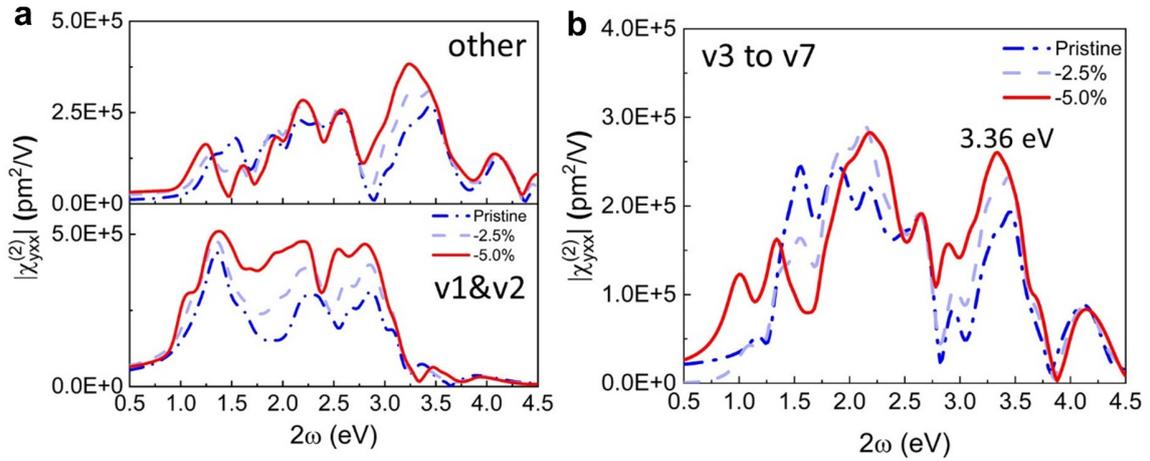

**Fig. S19.** Calculated $\left|\chi^{(2)}_{yyy}\right|$ for 2L A$\bar{\text{C}}$ RhI$_3$ under different pressures corresponding to different bands. **a.** The top two valence bands (v1 and v2) and other valence bands. **b.** Valence bands v3 to v7. V3 and v7 are the third and seventh top valence bands.

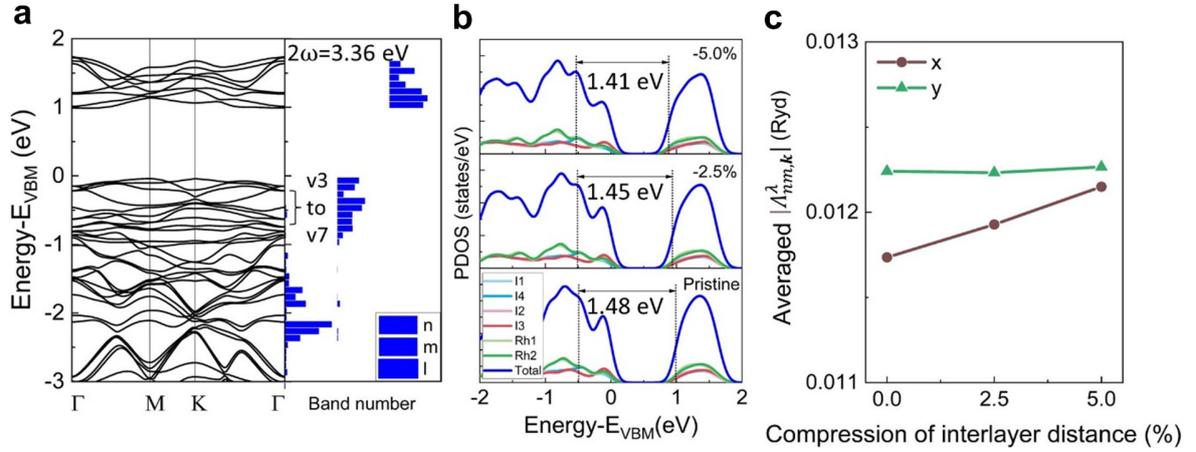

**Fig. S20. Analysis of SHG enhancement effects of $\chi^{(2)}_{yxx}$ at $2\omega = 3.36$ eV under pressure. a.** Band structure the statistical analysis of the bands corresponding to the transitions contribute dominantly to the real and imaginary part of SHG susceptibility element $\chi^{(2)}_{yxx}$ at 3.36 eV. Bands v3 to v7 are labelled. **b.** Partial density of states (PDOS) of 2L A$\bar{\text{C}}$ RhI$_3$ with and without a different compression of interlayer distance. The energy differences between the peak at ~0.5 eV and conduction band minimum are marked. **c.** Averaged absolute value of momentum matrix elements $|\Lambda^{\lambda}_{nm,\mathbf{k}}|$ ($n \in \{v1, v2\}, m \in \{CB\}, \lambda = x, y$) against the compression of interlayer distance.

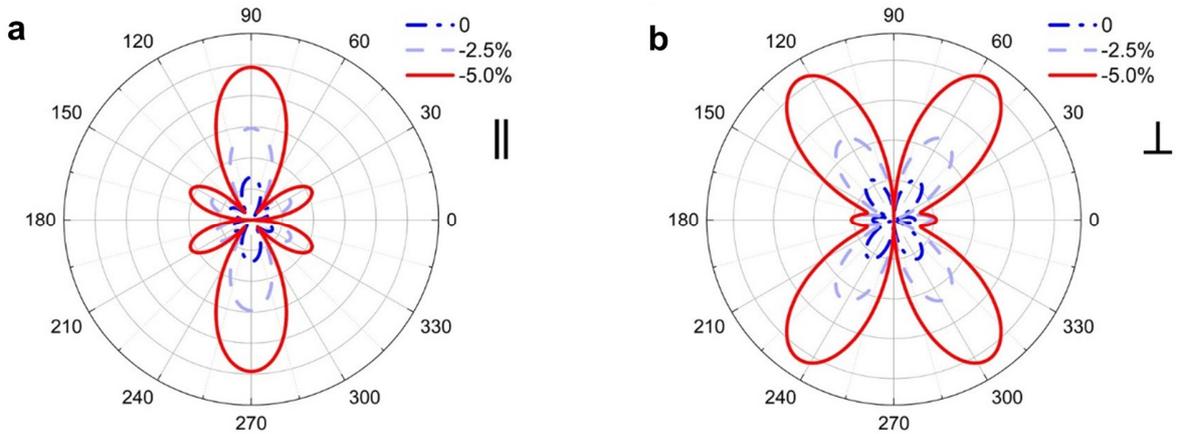

**Fig. S21. Polar plots of SHG response at 2.33 eV.** Calculated SHG intensity plotted as a function of polarization angle of the incident light for 2L A$\bar{\text{C}}$ RhI$_3$ in (**a**) parallel and (**b**) perpendicular configuration.

# Supplementary Tables

Table S1. Comparison of pressure-driven SHG intensity enhancement of various vdWs materials.

| Materials | SHG enhancement factor | Reference |
|---|---|---|
| $RhI_3$ | 6.9 (2.9 GPa) | This work |
| $NbOI_2$ | 3 (2.5 GPa) | Ref. [5] |
| $CuInP_2S_6$ | 10 (3.4 GPa) | Ref. [6] |
| InSe | No enhancement | Ref. [7] |
| $NbOCl_2$ | No enhancement | Ref. [8] |
| $BiOIO_3$ | No enhancement | Ref. [9] |

**References**

1. Seto, Y. & Ohtsuka, M. ReciPro: free and open-source multipurpose crystallographic software integrating a crystal model database and viewer, diffraction and microscopy simulators, and diffraction data analysis tools. *J Appl Cryst* **55**, 397–410 (2022).

2. Niu, Y. *et al.* Thickness-Dependent Differential Reflectance Spectra of Monolayer and Few-Layer $MoS_2$, $MoSe_2$, $WS_2$ and $WSe_2$. *Nanomaterials* **8**, 725 (2018).

3. Bing, D. *et al.* Optical contrast for identifying the thickness of two-dimensional materials. *Optics Communications* **406**, 128–138 (2018).

4. Li, H. *et al.* Rapid and Reliable Thickness Identification of Two-Dimensional Nanosheets Using Optical Microscopy. *ACS Nano* **7**, 10344–10353 (2013).

5. Fu, T. *et al.* Manipulating Peierls Distortion in van der Waals $NbOX_2$ Maximizes Second-Harmonic Generation. *J. Am. Chem. Soc.* **145**, 16828–16834 (2023).

6. Bu, K. *et al.* Enhanced Second-Harmonic Generation of van der Waals $CuInP_2S_6$ via Pressure-Regulated Cationic Displacement. *Chem. Mater.* **35**, 242–250 (2023).

7. Su, H. *et al.* Pressure-Controlled Structural Symmetry Transition in Layered InSe. *Laser Photonics Rev.* 7 (2019).

8. Ye, L. *et al.* Manipulation of nonlinear optical responses in layered ferroelectric niobium oxide dihalides. *Nat Commun* **14**, 5911 (2023).

9. Jiang, D. *et al.* Pressure-Driven Two-Step Second-Harmonic-Generation Switching in $BiOIO_3$. *Angew Chem Int Ed* **61**, (2022).